\documentclass[11pt]{article}
\usepackage{latexsym,comment,color}
\usepackage{amsmath,amsfonts,amsbsy,amssymb,hyperref}

 \csname
@addtoreset\endcsname{equation}{section}

\def\bec{\begin{center}}
\def\ec{\end{center}}
\def\a{\alpha} \def\ad{\dot{\a}} 

\def\b{\beta}  \def\bd{\dot{\b}} 
\def\c{\gamma} 
\def\C{\Gamma}
\def\d{\delta} 

\def\e{\epsilon} 
\def\ve{\varepsilon}

\def\k{\kappa}
\def\l{\lambda}
\def\L{\Lambda}
\def\m{\mu}
\def\n{\nu}

\def\s{\sigma}
\def\S{\Sigma}
\def\t{\tau}

\def\o{\omega}

\def\cO{{\cal O}}

\def\cE{{\cal E}}
\def\cM{{\cal M}}

\def\cV{{\cal V}}

\def\cU{{\cal U}}

\def\yb{{\bar y}}

\def\cO{{\cal O}}

\def\del{\partial}

\def\nn{\nonumber}
\newcommand{\eq}[1]{(\ref{#1})}

\newcommand{\w}[1]{\\[0.#1cm]}
\def\be{\begin{equation}}
\def\ee{\end{equation}}
\def\bea{\begin{eqnarray}}
\def\eea{\end{eqnarray}}
\def\ba{\begin{array}}
\def\ea{\end{array}}

\def\ft#1#2{{\textstyle{{\scriptstyle #1}
\over {\scriptstyle #2}}}}

\def\ad{\dot\alpha}
\def\bd{\dot\beta}

\def\Real{{\mathbb R}}
\def\Comp{{\mathbb C}}

\def\wt{\widetilde}
\def\wV{V}
\def\ty{\widetilde y}

\newcommand{\bse}{\begin{subequations}}
\newcommand{\ese}{\end{subequations}}

\begin{document}

%\begin{flushright}
% MIFP \\
% \vskip 8pt
%{\today}\\
%\end{flushright}

\vspace{10pt}

\begin{center}

%%%%%%%%%%%%%%%%%%%%%%%%%%%%%%%%%%%%%%%%%%%%%%%%%%%%%%%%%%%%%%%%%%%%

{\Large \textbf {On Exact Solutions and Perturbative Schemes\\ }}
\vspace{10pt}
{\Large\textbf {in Higher Spin Theory}}

%%%%%%%%%%%%%%%%%%%%%%%%%%%%%%%%%%%%%%%%%%%%%%%%%%%%%%%%%%%%%%%%%%%%

\vspace{15pt}

{\large Carlo Iazeolla$^1$, Ergin Sezgin$^{2}$ and Per Sundell$^{3}$}
\\[20pt]

\small{
\textit{ $^1$  NSR Physics Department, G. Marconi University \\  via Plinio 44, 00193 Rome, Italy} 
} 

\small{
\textit{   $^2$   Mitchell Institute for Fundamental Physics and Astronomy\\ Texas A\&M University
College Station, TX 77843, USA  }} 

\small{
\textit{ $^3$ Departamento de Ciencias F\'isicas, Universidad Andres Bello\\ Sazie 2212, Santiago de Chile}
}

\vspace{5pt}

%%%%%%%%%%%%%%%%%%%%%%%%%%%%%%%%%%%%%%%%%%%%%%%%%%%%%%%%%%%%%%%%%%%%

\vspace{15pt} {\large Abstract}

\end{center}

We review various methods for finding exact solutions of higher spin 
theory in four dimensions, and survey the known exact solutions of (non)minimal Vasiliev's equations. 
These include instanton-like and black hole-like solutions in (A)dS and Kleinian spacetimes. 
A perturbative construction of solutions with the symmetries of a 
domain wall is described as well.
Furthermore, we review two proposed perturbative schemes: one based on perturbative
treatment of the twistor space field equations followed by inverting
Fronsdal kinetic terms using standard Green's functions; and an alternative scheme 
based on solving the twistor space field equations exactly followed by introducing 
the spacetime dependence using perturbatively defined gauge functions.
Motivated by the need to provide a higher spin invariant characterization of the exact 
solutions, aspects of a proposal for a geometric description of Vasiliev's equation 
involving an infinite dimensional generalization of anti de Sitter space are revisited and improved.

\setcounter{page}{1}

\pagebreak

\tableofcontents

%%%%%%%%%%%%%%%%%%%%%%%%%%%%%%%%%%%%%%%%%%%%%%%%%%%%%%%%%%%%%%%%%%%%%
\section{Introduction}
%%%%%%%%%%%%%%%%%%%%%%%%%%%%%%%%%%%%%%%%%%%%%%%%%%%%%%%%%%%%%%%%%%%%%

Higher spin (HS) theory in four dimensions, in its simplest form and 
when expanded about its (anti-)de Sitter vacuum solution, describes 
a self-interacting infinite tower of massless particles of spin $s=0,2,4...$.
The full field equations, proposed long ago by Vasiliev \cite{vasiliev,Vasiliev:1990vu,Vas:more} (for reviews, see \cite{Vasiliev:1999ba,Didenko:2014dwa}), are a set
of Cartan integrable curvature constraints on master zero-, one- and 
two-forms living on an extension of spacetime by a non-commutative 
eight-dimensional twistor space.
The latter is fibered over a four-dimensional base, coordinated 
by a Grassmann-even $SL(2,\Comp)$-spinor oscillator $Z^A= (z^\alpha, \bar z^{\dot\alpha})$,
and the fiber is coordinatized by another oscillator 
$Y^A=(y^\alpha,\bar y^{\dot\alpha})$; the master fields are horizontal
forms on the resulting twelve-dimensional total space, valued an 
infinite-dimensional associative algebra generated by $Y^A$, that we 
shall denote by ${\cal A}$, and subject to boundary conditions on the 
base manifold.

A key feature of Vasiliev's equations is that they admit
asymptotically (anti-)de Sitter solution spaces, obtained 
by taking the HS algebra ${\cal A}$ to be an extension of the Weyl algebra, 
with its Moyal star product, by involutory chiral delta functions \cite{Iazeolla:2011cb,Iazeolla:2017vng}, 
referred to as inner Klein operators, relying on a realization of 
the star product using auxiliary integration variables \cite{Vasiliev:1999ba}.
Introducing a related class of forms in $Z$-space, that
facilitates a special vacuum two-form in twistor space,
the resulting linearized master fields can be brought to a 
special gauge, referred to as the Vasiliev gauge, in which 
their symbols defined in a certain normal order are real 
analytic in twistor space, and the master zero- and one-forms 
admit Taylor expansions in $Y$ at $Z=0$ in terms of Fronsdal 
fields on the mass shell and subject to physical boundary conditions.

%%%%%%%%%%%%%%%%%%%     Revision below   %%%%%%%%%%%%%%%%%%%%%%%%
%
Although Vasiliev equations take a compact and elegant form in the 
extended space, their analysis in spacetime proceeds in a weak field 
expansion which takes an increasingly complicated form beyond the leading order.
Indeed, they have been determined so far only up to quadratic order.  
In performing the weak field expansion, a number of challenges emerge. 
Firstly, obtaining these equations requires boundary conditions 
in twistor space, referring to the topology of $Z$ space and the 
classes of functions making up ${\cal A}$ 
\cite{Vasiliev:1990vu,ProkVas,Vasiliev:2015wma}. Proper way to 
pin down these aspects remains to be determined.
Second, the cosmological constant, $\Lambda$, which is necessarily
nonvanishing in Vasiliev's theory (as the transvection operators 
of the isometry algebra are realized in ${\cal A}$ as bilinears in $Y$), 
appears in the effective equations to its first power via critical mass 
terms, but also to arbitrary negative powers via non-local interactions 
\cite{Vasiliev:1999ba,Bekaert:2010hw}.
Thus, letting $\phi$ denote a generic Fronsdal field, it follows 
that $\partial \phi \sim \sqrt{|\Lambda|}\phi$ on-shell, 
and hence interactions with any number of derivatives are 
of equal relevance (at a fixed order in weak field amplitudes).
This raises the question of just how badly nonlocal are the HS field equations,
the attendant problem of divergences arising even at the level of the amplitudes 
\cite{Giombi:2009wh,Boulanger:2015ova} and what kind of field redefinitions are 
admissible. One guide available is the holographic construction of the bulk vertices \cite{Bekaert:2014cea,Bekaert:2015tva,Sleight:2016dba}. 
Clearly, it would be desirable to find the principles
that govern the nonlocal interactions, based on 
the combined boundary conditions in twistor space as well as 
spacetime, such that an order by order construction of the bulk vertices can proceed
from the analysis of Vasiliev equations. The simple and geometrical form of 
Vasiliev equations, in turn, may pave the way for the construction of an off-shell 
action that will facilitate the computation of the quantum effects.

In an alternative approach to the construction of HS equations in spacetime, 
it has been proposed to view Vasiliev's equations as describing stationary 
points of a topological field theory with 
a path integral measure based on a Frobenius--Chern--Simons 
bulk action in nine dimensions augmented by topological boundary terms,
which are permitted by the Batalin--Vilkovisky formalism, of which only
the latter contributes to the on-shell action \cite{Boulanger:2011dd,Boulanger:2015kfa}.
%
%%%%%%%%%%%%%%%%%%%%%%%%   end of revision   %%%%%%%%%%%%%%%%%%%%%%%
%

This approach combines the virtues of the on-shell approach
to amplitudes for massless particles flat spacetime with those of 
having a background independent action, in the sense that
the on-shell action is fixed essentially by gauge symmetries 
and given on closed form, which together with the background 
independence of Vasiliev's equations provides a machinery for
perturbative quantum computations around general backgrounds. 

In this context, it is clearly desirable to explore in more detail how
the choice of boundary conditions in the extended space influences the
classical moduli space of Vasiliev's equations, with the purpose of 
spelling out the resulting spaces, computing HS invariant
functionals on-shell, and examining how the strongly coupled spacetime 
nonlocalities are converted into physical amplitudes using the 
aforementioned auxiliary integral representation of star products 
in twistor space.

The aim of this article is to review three methods that have been used to 
find exact solutions of Vasiliev equations, and to describe two schemes 
for analyzing perturbations around them.  
In particular we will describe the gauge function method \cite{Vasiliev:1990bu,Bolotin:1999fa} for finding exact 
solutions and summarize the first such solution found in \cite{Sezgin:2005pv}, 
as well as its generalization to de Sitter spacetime studied in \cite{Iazeolla:2007wt} together with the solutions of for a chiral version of the theory with Kleinian $(2,2)$ signature.
As we shall see, this method uses the fact that the spacetime 
dependence of the master fields can be absorbed into gauge functions, 
upon which the problem of finding exact solutions 
is cast into relatively manageable deformed oscillator problem in twistor space.  
The role of different ordering schemes for star product 
as well as gauge choices to fix local symmetries in twistor 
space will also be discussed.  

Next, we will describe a refined gauge function method proposed 
in \cite{Iazeolla:2012nf}, where the twistor space equations are 
solved by employing separation of twistor variables and holomorphicity in the $Z$ space 
in a Weyl ordering scheme and enlarging the Weyl algebra in the 
fiber $Y$ space by inner Kleinian operators. 
This approach provides exact solution spaces in a particular gauge, 
that we refer to as the holomorphic gauge, after which the spacetime
dependence is introduced by means of a sequence of large gauge transformations, by 
first switching on a vacuum gauge function, taking the solutions to what 
we refer to as the $L$-gauge, where the configurations must be real 
analytic in $Z$ space, which provides an admissibility condition on
the initial data in holomorphic gauge. 
The solutions can then be mapped further to the 
Vasiliev gauge, where the linearized, or asymptotic, master fields, 
are real analytic in the full twistor space and obey a particular
gauge condition in $Z$ space which ensures that they consist of 
decoupled Fronsdal fields in a canonical basis; the required gauge
transformation, from the $L$ gauge to the Vasiliev gauge, can be
implemented in a perturbation scheme, which has so far been
implemented mainly at the leading order.
We will describe a black hole-like solution in some detail and 
mention other known solutions obtained by this method so far, 
including new solutions with six Killing symmetries \cite{cosmo}. 

We shall also outline a third method, in which the HS equations 
are directly tackled without employing gauge functions. 
In this method, solving the deformed oscillators
in twistor space also employs the projector formalism, 
though the computation of the gauge potentials does not rely 
on the gauge function method. 
The black hole-like solution found in this way in \cite{Didenko:2009td} 
will be summarized. 

We shall also review two approaches to the perturbative 
treatment of Vasiliev's equations. 
One of them, which we refer to as the normal ordered scheme, is 
based on a weak field expansion around (anti-)de Sitter 
spacetime \cite{Vas:more,Vasiliev:1999ba,Sezgin:2002ru}. 
It entails nested parametric integrals, introduced via a 
homotopy contraction of the de Rham differential in $Z$ space
used to solve the curvature constraints that have at least one 
form index in $Z$ space, followed by inserting 
the resulting perturbatively defined master fields
into the remaining curvature constrains with all form
indices in spacetime.
In an alternative scheme, the equations are instead solved 
exactly in at he aforementioned $L$-gauge, and  perturbatively 
realized large HS gauge transformation is then performed to achieve 
interpretation in terms of Fronsdal fields in asymptotically 
(anti-)de Sitter spacetimes in Vasiliev gauge \cite{Iazeolla:2017vng}.
The advantages of the latter approach in describing the fluctuations
around more general HS backgrounds will be explained.

A word of caution is in order concerning the usage of `black hole' 
terminology in describing certain types of exact solutions to HS equations. 
This terminology is, in fact, misleading in some respects since the 
notion of a line interval associated with a metric field is not 
HS invariant. 
Indeed, the apparent singular behaviour at the origin may in principle
be a gauge artifact. 
This point is discussed in more detail in Section 4.2. 
Moreover, given the nonlocal nature of the HS interactions, the 
formulation of causality, which is crucial in describing the 
horizon of a black hole, is a challenging problem without any 
proposal for a solution yet in sight; in fact, a more natural
physical interpretation of the black hole-like solutions may
turn out to be as smooth black-hole micro states \cite{Iazeolla:2017vng,Carlo}.
Another aspect of the known black hole-like solution in HS theory 
is that they activate fields of all possible spins, and apparently 
it is not possible to switch of all spins except one even in
the asymptotically AdS region.

So, what is meant by a black hole solution in HS theory? 
Firstly, the $SO(3)\times SO(2)$ symmetry of the solution 
(which is part of an infinite dimensional extended symmetry 
forming a subgroup of the HS symmetry group) is in common 
with the symmetry group arising in the asymptotically AdS BH 
solution of ordinary AdS gravity. 
Second, the solution contains a spin-two Weyl tensor field which takes 
the standard Petrov type D form, with a singularity at the origin;
more generally, the spin-s Weyl tensors are of a generalized 
Petrov type D form, given essentially by the s-fold direct 
products of a spin-one curvature of the Petrov type D form.
The BH terminology is thus used in the context of HS theory 
with the understanding that it is meant to convey these 
properties, albeit they do not constitute a rigorous 
definition of a black hole in HS theory.

The use of HS invariants for exact solutions to capture their 
physical characteristics has been considered and in some cases 
they have been computed successfully.
These particular invariants alone do not, however, furnish an 
answer to the question of whether it makes any sense to think
about event horizons in HS theory at all, and if so, how to define them;
in fact, their existence rather supports the aforementioned
micro state proposal, wherein the HS invariants
can be interpreted as extensive charges defining HS ensembles.

Motivated by the quest for giving a physical interpretation 
of the exact solution in the context of underlying HS symmetries, 
a geometrical approach to HS equations was proposed in \cite{Sezgin:2011hq}. 
We shall summarize this proposal in which the HS geometry 
is based on an identification of infinite dimensional structure 
group in a fibre bundle setting, and the related soldering phenomenon 
that leads to a HS covariant definition of classes of (non-unique) generalized vielbeins 
and related metrics, and as such an infinite dimensional generalization of $AdS$ geometry.
In doing so, we will improve the formulation of \cite{Sezgin:2011hq} by dispensing with the embedding of the relevant infinite dimensional coset space into a larger one that involves the extended HS algebra that includes the twistor space oscillators.

Finally, we are not aware of any exact solutions of HS theories in dimensions 
$D>4$ \cite{Vasiliev:2003ev,Bekaert:2005vh}, while in D=3, assuming that the 
scalar field is coupled to HS fields, we can refer to \cite{ProkVas,Didenko:2006zd,Iazeolla:2015tca} 
for the known solutions. Purely topological HS theory, which has no dynamical degrees of 
freedom, and which allows a more rigorous definition of black holes, is known to 
admit many exact solutions whose description goes beyond the scope of this review. 

%%%%%%%%%%%%%%%%%%%%%%%%%%%%%%%%%%%%%%%%%%%%%%%%%%%%%%%%%%%%%%%%%%
\section{Vasiliev Equations}
%%%%%%%%%%%%%%%%%%%%%%%%%%%%%%%%%%%%%%%%%%%%%%%%%%%%%%%%%%%%%%%%%%

\subsection{Bosonic model in $(A)dS$}
%%%%%%%%%%%%%%%%%%%%%%%%%%%%%%%%%%%%%%%%%%%%

Vasiliev's theory is formulated in terms of horizontal forms on a non-commutative fibered 
space ${\cal C}$ with four-dimensional non-commutative symplectic 
fibers and eight-dimensional base manifold equipped
with a non-commutative differential Poisson structure.
On the total space, the differential form algebra $\Omega({\cal C})$ is
assumed to be equipped with an associative degree preserving 
product $\star$, a differential $d$, and an Hermitian conjugation 
operation $\dagger$, that are assumed to be mutually compatible. 
The base manifold is assumed to be the direct product of a 
commuting real four-manifold ${\cal X}_4$ with coordinates $x^\mu$,
and a non-commutative real four-manifold ${\cal Z}_4$ with coordinates
$Z^A$; the fiber space and its coordinates are denoted by ${\cal Y}_4$
and $Y^A$, respectively.
The non-commutative coordinates are assumed to obey
\be 
[Y^A,Y^B]_\star=2iC^{AB}\ ,\qquad 
[Z^A,Z^B]_\star=-2iC^{AB}\ ,\qquad
[Y^A,Z^B]_\star=0\ ,
\ee
where $C_{AB}$ is a real constant antisymmetric matrix. The non-commutative space is 
furthermore assumed to have a compatible complex structure, such that 
\be 
Y^A=(y^{\alpha},\bar y^{\dot\alpha})\ ,\qquad
Z^A=(z^{\alpha},\bar z^{\dot\alpha})\ ,
\ee
\be 
(y^{\alpha})^\dagger=\bar y^{\dot\alpha}\ ,\qquad
(z^{\alpha})^\dagger=-\bar z^{\dot\alpha}\ ,
\ee
where the complex doublets obey
$[y^{\alpha},y^{\beta}]_\star=2i \epsilon^{\alpha\beta}$ and 
$[z^{\alpha},z^{\beta}]_\star=-2i \epsilon^{\alpha\beta}.$
The horizontal forms can be represented as sets of
locally defined forms on ${\cal X}_4\times {\cal Z}_4$ valued 
in oscillator algebras ${\cal A}({\cal Y}_4)$ generated by the fiber 
coordinates glued together by transition functions, that we shall
assume are defined locally on ${\cal X}_4$, resulting in a bundle 
over ${\cal X}_4$ with fibers given by  $\Omega({\cal Z}_4)\otimes {\cal A}({\cal Y}_4)$.
The algebra ${\cal A}({\cal Y}_4)$ can be given in various bases; we shall use 
the Weyl ordered basis,  and the normal ordered basis consisting of 
monomials in $a_\pm= Y\pm Z$ with  $a_+$ and $a_-$ oscillators standing to 
the left and right, respectively.
We assume that the elements in $\Omega({\cal Z}_4)\otimes {\cal A}({\cal Y}_4)$
have well-defined symbols in both Weyl and normal order.   The normal order reduces to 
Weyl order for elements that are independent of either $Y$ or $Z$, and in the cases where 
depend  on both $Y$ and $Z$, they can be composed using the Fourier transformed  twisted 
convolution formula in normal ordered scheme as
\be
(f\star g)(Z;Y) =
\frac{1}{(2\pi)^4} \int_{{\Real^4}\times {\Real^4}}  d^4U d^4V
f(Z+U;Y+U)\, g(Z-V; Y+V)\, e^{i V^A U_A}\ .
\label{star}
\ee

The model is formulated  in terms of a zero-form $\Phi$, a one-form 
\be
A=dx^\mu W_\mu + dz^\alpha V_\alpha + d\bar z^{\dot\alpha} \bar V_{\dot\alpha}\ ,
\ee
and a non-dynamical  holomorphic two-form 
\be
J:=-\frac{ib}4 dz^\alpha \wedge dz_\alpha\, \kappa\ ,
\ee
with Hermitian conjugate $\overline J=(J)^\dagger$, where $b$ is a complex parameter and 
\be 
\kappa:=\kappa_y \star \kappa_z\ ,\qquad 
\kappa_y:=2\pi \delta^2(y)\ ,\qquad 
\kappa_z:=2\pi \delta^2(z)\ ,\ee
are inner Klein operators obeying 
$\kappa_y\star f \star \kappa_y=\pi_y(f)$ and $\kappa_z\star f \star \kappa_z=\pi_z(f)$
for any zero-form $f$, where  $\pi_y$ and $\pi_z$ are the
automorphisms of $\Omega({\cal Z}_4)\otimes {\cal A}({\cal Y}_4)$ defined
in Weyl order by
\be
\pi_y (x;z,\bar z; y,\bar y) = (x;z,\bar z;- y,\bar y)\ ,\qquad 
\pi_z (x;z,\bar z; y,\bar y) = (x;-z,\bar z;y,\bar y)\ .
\ee
It follows that $dJ=0,\ J\star f=\pi(f)\star J$ and $\pi(J)=J$, \emph{idem} $\overline J$, with 
\be
 \pi: =\pi_y \circ \pi_z\ , \qquad \bar\pi: =\pi_{\bar y}\circ \pi_{\bar z} \ .
 \ee

It is useful to note that the inner Kleinian takes the following forms in different ordering schemes:
\be
\kappa = \begin{cases} e^{iy^\a z_\a} \qquad\qquad\qquad\  \mbox{in normal ordering scheme}\\ 
(2\pi)^2 \delta^2(y) \delta^2 (z)  \qquad \mbox{in Weyl ordering scheme} \end{cases}
\ee

The nonminimal and minimal models with all integer spins and only even spins, respectively, are obtained by imposing the conditions
\bea 
\mbox{Non-minimal model ($s=0,1,2,3,...$)}&:& \pi \circ \bar\pi(A) = A\ ,\qquad \pi \circ \bar\pi(B) =B\ ,
\label{nonminmod}\\[5pt]
\mbox{Minimal model ($s=0,2,4,...$)}&:& \tau(A) =
-A\
,\qquad \tau(B)\ =\ \bar\pi(B)\ ,
\label{minmod}
\eea
where $\t$ is the anti-automorphism
\be 
\tau( x^\mu; Y^A, Z^A)  =  f(x^\mu; iY^A, -iZ^A)\ ,\qquad \t( f\star g)\ =\ \t(g)\star\t(f)\ ,
\label{tau}
\ee
It follows that $\tau(J, \bar J)= (-J,-\bar J)$. 
Models in Lorentzian spacetimes with cosmological
constants $\Lambda$ are obtained by imposing reality conditions as follows  \cite{Iazeolla:2007wt}:
\be
\rho(B^\dagger)=\pi(B)\ ,\qquad \rho(A^\dagger)=-A\ ,\qquad
\rho:=\left\{ \begin{array}{ll} \pi\ ,&\Lambda>0 \\
{\rm Id}\ ,&\Lambda<0\end{array}\right.
\ee

Basic building blocks for Vasiliev equations are the curvature and twisted-adjoint covariant derivative defined by
\be
F :=dA+A\star A\ ,\qquad DB :=dB+[A,B]_\pi
\ ,
\label{curvatures}
\ee
respectively, where the $\pi$-twisted star commutators is defined as 
\be
[f,g]_\pi:=f\star g-(-1)^{|f||g|}g\star \pi(f)\ ,
\ee
and
\be
d:= d_x + d_Z\ ,\qquad d_x= dx^\mu\partial_\mu\ ,\qquad d_Z= dz^\a \partial_\a + d\bar z^{\ad}\bar \partial_{\ad}\ .
\ee
Vasiliev equations of motion are given by 
\be
F+ B \star (J- {\overline J})=0\ ,\qquad DB=0\ ,
\label{ve}
\ee
which are compatible with the kinematic 
conditions and the Bianchi identities,
implying that the classical solution space 
is invariant under the following infinitesimal 
gauge transformations:
\be 
\delta A=D\e:=d\e+[A,\e]_\star\ ,\qquad \delta B = -[\e,B]_\pi\ ,
\ee
for parameters obeying the same kinematic conditions as the connection. 

It remains to be a challenging problem to determine if these equation can be derived for a suitable tensionless, or critical tension, limit followed by a consistent truncation of string field theory on a background involving $AdS_4$. It will be very interesting to also determine if this equations follow from a consistent quantization of a topological string field theory. For a more detailed discussion and progress in this direction, see \cite{Engquist:2005yt,Arias:2016agc}. The component fields of $V_A$ do not 
transform properly under the Lorentz transformations generated by 
$(\frac1{4i}(y^\a y^\b-z^\a z^\b)-h.c)$. To remedy this problem and achieve manifest Lorentz covariance,  one introduces the field-dependent Lorentz generators \cite{Vasiliev:1999ba,Sezgin:2002ru}
\be
M_{\a\b}= M^{(0)}_{\a\b} +S_{(\a}\star S_{\b)} \ ,\qquad M^{(0)}_{\a\b} := y_{(\a}\star y_{\b)}-z_{(\a}\star z_{\b)}\ ,
\label{LT}
\ee
and their complex conjugates, where 
\be 
S_\a  =  z_\a-2i V_\a\ ,\qquad \bar S_{\dot\a}=\bar z_{\dot\a}-2i\bar V_{\dot\a}\ .
\label{defS}
\ee
Next one defines
\be
W'_\mu=W_\mu - \frac{1}{4i}\left(\omega_\mu^{\a\b} M_{\a\b}
+\bar\omega_\mu^{\dot\a\dot\b} \bar M_{\dot\a\dot\b}\right)\ ,
\label{WP}
\ee
where $(\omega_\mu^{\a\b},\omega_\mu^{\dot\a\dot\b})$ is the
canonical Lorentz connection.  
It  is defined up to tensorial shifts \cite{Sezgin:2011hq} 
that can be fixed by requiring that the projection of $W'$ onto 
$M^{(0)}_{\a\b}$ and its complex conjugate, vanish at $Z=0$, that is
\be
\frac{\partial^2}{\partial y^\a\partial y^\b} W'\vert_{Y=Z=0} =0 \ , \qquad
\frac{\partial^2}{\partial \bar y^{\ad}\partial \bar y^{\bd}} W'\vert_{Y=Z=0} =0 \ .
\label{ss}
\ee
The above redefinitions ensure that under the Lorentz transformations with parameters 
\be
\e_L = \frac{1}{4i} \Lambda^{\a\b} M_{\a\b}\ , \qquad \e^{(0)}_L = \frac{1}{4i} \Lambda^{\a\b} M^{(0)}_{\a\b}\ ,
\ee
the master fields transform properly under the Lorentz transformations as \cite{Sezgin:2002ru}
\bea
\delta_L B &=& [\e^{(0)}_L,B]_\star\ ,
\w2
\delta_L S_\a &=& [\e^{(0)}_L,S_\a]_\star + \Lambda_\a{}^\b\, S_\b\ ,
\qquad {\rm idem}\quad  \ \ \bar S_{\ad}\ ,
\w2
\delta_L W'_\mu &=& [\e^{(0)}_L,W'_\mu]_\star + \frac{1}{4i} \left(\partial_\mu \Lambda^{\a\b} M_{\a\b} +h.c.\right) 
\eea
Using \eq{defS}, the component form of  Vasiliev equations take the form
\bea 
&& d_x W + W\star W = 0\ ,
\label{B1}\w2
&& d_x B- [W, B]_\pi = 0\ ,
\label{B2}\w2
&& d_x S_\a+[W,S_\a]_\star = 0\ ,\qquad  d_x\bar S_{\ad}+[W,\bar S_{\ad}]_\star =  0\ ,
\label{S1}\w2
 && [S_\a,B]_\pi =0\ , \qquad [\bar S_{\ad},B]_\pi =0\ ,
\quad [S_\a, \bar S_{\bd}]_\star=0\ ,
\label{S2}\w2
&& [S^\a,S_\a]_\star =4i(1- b B\star\kappa)\ ,\quad[ \bar S^{\ad}, \bar S_{\ad}]_\star\ =\  4i(1- \bar b B\star\bar\kappa)\ .
\label{S3}
\eea
This is the form of the equations typically used to seek exact solutions as it displays the role of the 
deformed oscillator algebra in the last two equations. Here one may exploit the technology developed in study of noncommutative field theories and construction of projection operators in a suitably defined oscillator space. 

%%%%%%%%%%%%%%%%%%%%%%%%%%%%%%%%%%%%%%%%%%%%%%%%%%%%%%%%%%%%%
\subsection{The nonminimal chiral model in Kleinian space}
%%%%%%%%%%%%%%%%%%%%%%%%%%%%%%%%%%%%%%%%%%%%%%%%%%%%%%%%%%%%%

In this model the spinor oscillators are now representations of $SL(2,\Real)_L\times SL(2,\Real)_R$, and as such their hermitian conjugates are now given by  
\be
(y^\a)^\dagger =  y^{\a}\ ,\quad (z^\a)^\dagger = -z^\a\ ,
\quad  (\bar y^{\ad})^\dagger = \bar y^{\ad}\ ,\quad (\bar z^{\ad})^\dagger = 
-\bar z^{\ad} \ .
\label{kr}
\ee
The field equations are now given by
\be
F= J\star B \ , \qquad DB=0\ ,\qquad     J:=-\frac{i}4 dz^\alpha \wedge dz_\alpha\, \kappa\ ,
\ee
with reality conditions $A^\dagger = -A$ and $B^\dagger = \pi(B)$,  and kinematical conditions
\bea 
\mbox{Minimal model ($s=0,2,4,...$)}&:& \tau(A) =
- A\
,\qquad \tau(B)\ =\ \bar\pi(B)\ ,
\label{minmod}\\[5pt]
\mbox{Non-minimal model ($s=0,1,2,3,...$)}&:& \pi \circ \bar\pi(A) = A\ ,\qquad \pi \circ \bar\pi(B) =B\ .
\label{nonminmod}
\eea
The field equations in components take the same form as in \eq{B1}-\eq{S3}, but now
with
\be b=1\ ,\qquad  \bar b=0\ .\ee
These models are referred to as being chiral in view of the half-flatness condition on the twistor space curvature, namely $F_{\dot\a\dot\b}=0$. These models admit the coset space
$H_{3,2}= SO(3,2)/SO(2,2)$ as a vacuum solution, which has the Kleinian signature $(2,2)$.
For a detailed  description of these spaces, including the curved Kleinian geometries, see \cite{Barrett:1993yn}. Our motivation for highlighting this case is due to its being the first exact solution of Vasiliev equation in which all HS fields are nonvanishing, and that the Kleinian geometry is relevant to $N=2$ superstring as well as integrable models.

%%%%%%%%%%%%%%%%%%%%%%%%%%%%%%%%%%%%%%%%%%%%%%%%%%%%%%%%%%%%%%%%%%%%%%%%%
\section{Gauge Function Method and Solutions}\label{sec:zspace}
%%%%%%%%%%%%%%%%%%%%%%%%%%%%%%%%%%%%%%%%%%%%%%%%%%%%%%%%%%%%%%%%%%%%%%%%%

\subsection{The method}
%%%%%%%%%%%%%%%%%%%%%%%%%%%%%%%%%

In order to construct solutions to Vasiliev's equations, one may 
consider the approach \cite{Bolotin:1999fa} in which they are
homotopy contracted in simply connected spacetime regions $U$
to deformed oscillator algebras in twistor space at a base point 
$p\in U$; the constraints 
 \be
 F_{\m\n}\ =\ 0\ ,\qquad F_{\m\a}\ =\ 0\
,\qquad D_\m B  = 0\ ,
\label{xsp2}
 \ee
are thus integrated in $U$ using a gauge function $g=g(Z,Y|x)$ 
obeying
\be 
g|_{p} = 1\ ,
\ee
and initial data
\be
 B' = B|_{p}\ ,\qquad  S'_A\ = S_A|_{p}\ ,
 \label{phiprime}
 \ee
subject to
\bse
\bea
&& [S'_\a,B']_\pi =0\ , \qquad [\bar S'_{\ad},B']_\pi =0\ ,
\qquad [S'_\a, \bar S'_{\bd}]_\star=0\ ,
\label{P1}\w2
&& [S'^\a,S'_\a]_\star = 4i(1- b B'\star\kappa)\ ,
\quad[ \bar S'^{\ad}, \bar S'_{\ad}]_\star =  4i(1- \bar b B'\star\bar\kappa)\ .
\label{P2}
\eea
\label{P12}
\ese
The fields in $U$ can then be expressed explicitly as
\be
W_\mu = g^{-1}\star \partial_\mu g\ ,
\qquad 
S_A = g^{-1}\star S'_A \star g\ ,
\qquad 
B= g^{-1}\star B'\star \pi(g)\ ,
\label{Leq}
 \ee
after which the Lorentz covariant HS gauge fields can be obtained from
\eq{WP} subject to \eq{ss}, which serves to determine the spin connection 
$\omega_{\mu ab}$. 
Thus, the deviations in the spacetime HS gauge fields away from the topological
vacuum solution, that is the solution with $W_\mu=0$, thus come from the gauge function $g$ as well as the non-linear shift on the account of achieving manifest Lorentz invariance.
The deformed oscillator algebra requires a choice of
topology for ${\cal Z}_4$, initial data for $B'$ and 
a flat background connection.
In what follows, we shall assume that ${\cal Z}_4$ has the
topology of $\Real^4$ with suitable fall-off conditions at
infinity \cite{Boulanger:2015kfa,Iazeolla:2017vng}, and impose 
\be
C'(Y)\ = B'|_{Z=0}\ ,\qquad S'_A\vert_{C'=0}=Z_A ;
\ee
for nontrivial flat connections on ${\cal Z}_4$, that are not pure gauge,
see \cite{Iazeolla:2007wt}.
The gauge function represents a gauge transformation that is large in
the sense that it affects the asymptotics of gauge fields so as to 
introduce additional physical degrees of freedom to the system,
over and above those contained in the twistor space initial data and 
flat connection; strictly speaking, in order to define such transformations,
one should first introduce a set of classical observables forming a BRST 
cohomology modulo a set of boundary conditions on ghosts, after which
a large gauge transformation is a gauge transformation that does not 
preserve all the classical observables.
In particular, in order to describe asymptotically maximally symmetric, 
or Weyl flat, solutions, one may take
\be g|_{B'=0}=L\ ,\ee
where $L=L(Y|x)$ is a metric vacuum gauge function, to be described 
below.
In order to obtain exact solutions, we shall choose $g=L$ for 
all $C'$, that we refer to as the $L$-gauge.
However, in order to extract Fronsdal fields in the asymptotic 
region, one has to impose a gauge condition in twistor space 
to the leading order in the weak field expansion in the asymptotic 
region, which introduces a dressing of the vacuum gauge function
by an additional perturbatively determined gauge function; 
see Section 6.
%

%%%%%%%%%%%%%%%%%%%%%%%%%%%%%%%%%%%%%%%%%%%%%
\subsection{Vacuum solutions}
%%%%%%%%%%%%%%%%%%%%%%%%%%%%%%%%%%%%%%%%%%%%%

In order to obtain solutions containing locally maximally symmetric
asymptotic regions, one may take the gauge function $L(Y|x)$ to be 
corresponding coset representatives.
In what follows, we shall focus on the spaces
$AdS_4=SO(3,2)/SO(3,1)$, $dS_4=SO(4,1)/SO(3,1)$ and 
$H_{3,2}:=SO(3,2)/SO(2,2)$, which can be realized
as the embeddings
\be
X^A X^B \eta_{AB} \equiv -(X^0)^2 +(X^1)^2 + (X^2)^2  + (X^3)^2  +\e (X^5)^2= -\lambda^{-2}\ ,
\label{emb1}
\ee
where
\be
\left(\e, \frac{\lambda^2}{|\lambda^2|}\right) = 
\begin{cases} 
(-,+)\ \ {\rm for} \ \ AdS_4\\ 
(+,-)\ \ {\rm for}\ \  dS_4\\ 
(-,-)\ \ {\rm for}\ \  H_{3,2}
\end{cases}
\label{3s}
\ee 
These spaces can be conveniently described in a unified fashion
using the stereographic coordinates $x^a_\pm (a=0,1,2,3)$ obtained 
by means of the parametrization
\bea
&& X^M|_{U_\pm} \approx \left(\frac{2x^a_\pm}{1-\lambda^2 x^2_\pm},
\pm \ell\frac{1+\lambda^2 x^2_\pm}{1-\lambda^2 x^2_\pm}\right)\ ,\qquad
-1 \le \lambda^2 x^2_\pm <1\ ,\w2
&& x_\pm^2 := x^a_\pm x^b_\pm \eta_{ab}\ , 
\qquad \eta_{ab} = {\rm diag} (\e, \lambda^2/|\lambda^2|, +,+)\ , \qquad \ell =|\lambda|^{-1}\ ,
\label{sc}
\eea
where $U_\pm$ denotes the two stereographic coordinates charts,
each covering one half of the space \eq{3s}; on the
overlap one has $\lambda^2x^2_\pm=-1$, and the coordinate 
transition function 
\be 
x^a_\pm = R^a(x_\mp)\ ,\qquad \lambda^2x_\pm^2<0\ ,
\ee 
where the reflection map 
\be 
R^a(v):=-\frac{v^a}{\lambda^2 v^2}\ .
\ee
The boundary is given by $\l^2x^2_\pm=1$,
which has the topology of $S^2\times S^1$
in the case of $AdS_4$ and $H_{3,2}$, and 
$S^3\cup S^3$ in the case of $dS_4$.
Instead of covering the vacuum manifold 
with two charts, one may extend either one of
the charts to $\Real^4\setminus \{ x^a:\lambda^2 x^2=1\}$, 
which provides a global cover using a single chart, 
with the understanding that $\{ x^a:\lambda^2 x^2=1^-\}\cup 
\{ x^a:\lambda^2 x^2=1^+\}$ provides a two-sheeted
cover of the boundary.
The induced line element $ds_0^2 = dX^A dX^B \eta_{AB}|_{ \lambda^2 X^2=-1}$ is given by
\be 
ds^2_0 = {4 dx^2\over (1-\lambda^2x^2)^2}\ .
\label{gm}
\ee
On $|\lambda^2| x^2<1$, the corresponding vacuum gauge function
\bea 
L &=& {2h\over 1+h} \exp (-iy a \yb)\ , 
\label{L}
\eea
where
\be
\label{d1}
a_{\a\ad}={\lambda x_{\a\ad}\over 1+h}\ ,\qquad
x_{\a\ad}=(\s^a)_{\a\ad} x_a\ ,\qquad h =  \sqrt{1-\lambda^2x^2}\ .
\ee
\be
W_0 \equiv  e_0+\omega_0=L^{-1}\star dL
= \frac{1}{4i} \left[ \omega_0{}^{\a\b} y_\a y_\b +\bar \omega_0{}^{\ad\bd} \bar y_{\ad} \bar y_{\bd} + 2 e_0{}^{\a\ad} y_\a \bar y_{\ad} \right]\ ,
\label{ads4}
\ee
where  
\be 
e_0{}^{\a\ad} = -{\l(\s^a)^{\a\ad}dx_a\over h^2}\ ,\qquad 
\omega_0{}^{\a\b} = - {\l^2(\s^{ab})^{\a\b} dx_a x_b\over h^2}\ .
\label{adseo}
\ee
A global description can be obtained using two gauge functions
$L_\pm=L(Y|x_\pm)$ defined on $U_\pm$; the $Z_2$-symmetry implies 
that if $\Phi_\pm|_{p_\pm}=C'$, where $p_\pm:=x_\pm^{-1}(0)$,
then the two locally defined solutions on $U_\pm$ can be 
glued together using the gauge transition function 
$T^+_-:= L^{-1}_+\star L_-=1$ defined on the overlap region
where $\lambda^2 x_\pm^2=-1$.

For later purposes, it is convenient to introduce alternative coordinate systems which 
are defined by the embeddings (with $|\lambda|^2=1$)

\be
\begin{split}\label{ads}
AdS_4:  \quad x^2>0:\quad & X^0 = \sinh\tau \sinh\psi\ , \ \  X^i= n^i\cosh\tau \sinh\psi\ ,
\ \  X^5= \cosh\psi\ ,
\w2
\quad\quad\qquad x^2<0: \quad & X^0 = \sin\tau \cosh\psi\ ,  \ \  X^i= n^i\sin\tau \sinh\psi\ ,
 \ \  X^5= \cos\tau\ ,
\end{split}
\ee
\be
\begin{split}\label{ds}
dS_4: \   \quad x^2>0: \quad & X^0 = \sinh\tau \sin\psi\ ,\ \ X^i= n^i\cosh\tau \sin\psi\ , \ \ X^5= \cos\psi\ ,
\w2
\quad\quad\qquad x^2<0: \quad & X^0 = \sinh\tau \cosh\psi\ ,\  X^i= n^i\sinh\tau \sinh\psi\ ,
\ X^5= \cosh\tau\ ,
\end{split}
\ee
The metrics for (A)dS in these coordinate systems are given in \eq{adsm}-\eq{dsm}. In the case of $H_{3,2}$, we will find the following coordinate system to be useful (with $|\lambda^2|=1$)
\be
X^0 = r \sin t\ ,\qquad X^i = n^i \sqrt{1+r^2}\ ,\qquad  X_5 = r \cos t\ ,
\label{kcs}
\ee
where $n^i n^j \delta_{ij}=1$, $0\le r\le \infty$ and $0\le t \le 2\pi$. 

%%%%%%%%%%%%%%%%%%%%%%%%%%%%%%%%%%%%%%%%%%%%%%%%%%%%%%%%%%%%%%%%%%%%%%%%%%%%%%%
\subsection{Instanton solutions of minimal model in (anti) de Sitter space}
%%%%%%%%%%%%%%%%%%%%%%%%%%%%%%%%%%%%%%%%%%%%%%%%%%%%%%%%%%%%%%%%%%%%%%%%%%%%%%%

Having obtained a vacuum gauge function, the next task is to solve the deformed 
oscillator problem \eq{P12} subject to the initial data in twistor space.
To this end, it is helpful to constrain the primed configurations further by
assuming that they preserve a nontrivial amount of HS symmetries.
This can be achieved by imposing symmetry conditions by seeking a subspace $\mathfrak k'$ of HS gauge parameters  $\e'$ that are unbroken, \emph{i.e.}
\be 
\delta_{\e'} B'=0\ ,\qquad \delta_{\e'} S'_\alpha=0\ ,
\label{gsc}
\ee
for all $\e'\in\mathfrak k'$; upon switching on the gauge function
$g$, the resulting full solution is
invariant under gauge parameters in the space 
$\mathfrak{k}=g^{-1}\star \mathfrak k'\star g$.
For example, one may require that an $n$-dimensional subalgebra 
$\mathfrak g_n$ of the maximal finite dimensional subalgebra 
$\mathfrak{g}_{10}$ of the HS algebra remains unbroken, which 
implies that $\mathfrak k'$ is given by the intersection of 
${\rm Env}(\mathfrak{g}_n)$ and the HS algebra. 
In particular, taking $n=10$ yields the vacuum solution 
\be
W'= W_0\ , \qquad  S_A=z_A\ ,\qquad  B=0 \ ,
\label{vac}
\ee
which preserves the HS algebra itself.
Taking $n<10$, the first distinct cases with nontrivial 
Weyl zero-form arise for $n=6$; the space
$\mathfrak k'$ is then given exactly, as we shall describe below for a particular realization of ${\mathfrak g}_6$, or perturbatively. In the latter case the ${\mathfrak g}_6$ will be realized as an algebra of the  HS algebra in the leading order.

In \cite{Sezgin:2005pv}, asymptotically anti-de Sitter solutions with 
$\mathfrak{g}_6=\mathfrak{o}(1,3)$ were constructed by taking 
$\mathfrak{k}$ to be generated by the full Lorentz generators 
$M_{\a\b}$ from \eq{LT}.
Thus, in the primed basis, the corresponding symmetry conditions read
\bea
&& [M'_{\a\b},B']_\pi = 0\ , \qquad [M'_{\a\b}, S'_\gamma]_\star =0
\ , \qquad [M'_{\a\b}, \bar S'_{\dot\gamma}]_\star =0\ ,
\label{inv2}
\eea
and complex conjugates, where 
$M'_{\alpha\beta}=M^{(0)}_{\alpha\beta}+S'_{(\alpha}\star S'_{\beta)}$,
which are given by $y_{\a}y_\b$ plus perturbative corrections,
and whose star product commutators close modulo Lorentz transformations acting 
on the component fields; thus, consistency of the invariance conditions 
implies that all canonical Lorentz tensors that are not singlets must vanish.
Alternatively, it is possible to use other embeddings of $\mathfrak{o}(1,3)$ into the algebra
of primed HS gauge transformations; for example, one can simply take $y_{\a}y_\b$,
as we shall comment on in Section 5, though it remains an open problem whether 
the resulting solutions are gauge equivalent to those that will be presented below.
Taking $\mathfrak{g}_n$ to be generated by unperturbed functions of $Y$
is useful, however, in considering unbroken symmetry algebra involving transvection 
operators, as we shall spell out in further detail in Section 5 in the case 
of domain walls and related time dependent solutions.

Turning to \eq{inv2}, the simplest possible ansatz for $B'$ is a constant,
\emph{viz.}
\be B'=\nu\ ,\label{intsol1}\ee
which leads to a deformed oscillator problem with exact solutions closely
related to those of the 3D HS theory constructed by Prokushkin and Vasiliev \cite{ProkVas}.
Adapting to the 4D Type A model, for which $b=1$, the following solution for 
the twistor space connection was found in \cite{Sezgin:2005pv}
\bea
S'_\a &=& z_\a + z_\a \int_{-1}^1 dt~ q(t)\, \exp \left( \frac{i}2(1+t)\,u \right)\ ,
\qquad u:= y^\a z_\a\ ,
\label{tcp} \\[4pt] 
 q(t) &=& -{\nu\over 4}\left({}_1\!
F_1\left[\frac12;2;{\nu\over 2}\log \frac 1{t^2}\right]+t\,{}_1\!
F_1\left[\frac12;2;-{\nu\over 2}\log \frac 1{t^2}\right]\right)\ .
\label{intconn}
\eea
Expanding $\exp (it u /2)$ results in integrals of the
degenerate hypergeometric functions times positive algebraic powers
of $t$, which improve the convergence at $t=0$. 
Thus $V_\a$ is a power series expansion in $u$ 
with coefficients that are functions of $\nu$ that are 
well-behaved provided this is the case for the coefficient 
of $u^0$. 
This is the case for $\nu$ in some
finite region around $\nu=0$, as discussed in detail in \cite{Sezgin:2005pv,Sezgin:2005hf}. 
Indeed, as we shall see below after carrying out the integration over $t$, $\nu$ must lie in the interval $-3\le \nu \le 1.$

The solution in spacetime is obtained in the two regions $\l^2x^2<1$
and $\l^2\widetilde x^2<1$ using the stereographic gauge functions 
$L\equiv L(x|Y)$ and $\widetilde L\equiv L(\tilde x|Y)$ where $x$ and $\tilde x$
are related by the reflection map in the overlap region where
$\l^2x^2<0$ and $\l^2\widetilde x^2<0$. 
From \eq{d1}, one finds \cite{Sezgin:2005pv}
\be 
B = \nu (1-\lambda^2x^2)\exp\left[-i\l x^{\a\ad}y_\a \yb_{\ad}\right]\ .
\label{wphi}
\ee
This shows that the physical scalar field is given in the
$x^a$-coordinate chart by
\be 
\phi(x) = B|_{Y=Z=0} = \nu (1-\lambda^2x^2)\ ,\qquad \l^2x^2<1\ ,
\label{scalarsol}
\ee
while the Weyl tensors for spin $s=2,4,\dots$ vanish. Using instead $\widetilde L$, 
the physical scalar field in the $\widetilde x^a$-coordinate chart is given by
\be \widetilde \phi\ =\ \nu(1-\lambda^2\widetilde x^2)\ ,\qquad
\l^2\widetilde x^2<1\ .\ee
As a result, the two scalar fields are related by a duality
transformation in the overlap region
\be \widetilde\phi(\tilde x)\ =\ {\nu\phi(x)\over \phi(x)-\nu}\
,\qquad \lambda^2 x^2=(\lambda^2\widetilde x^2)^{-1}<0\
.\label{duality}\ee
Thus, if the transition takes place at $\lambda^2
x^2=\lambda^2\widetilde x^2=-1$, then the amplitude of the
physical scalar never exceeds $2\nu$.

The master fields $S_A$ and $W'_\mu$ are obtained from \eq{Leq} and \eq{WP}.
The generating functional for the spacetime gauge fields is given by 
\cite{Sezgin:2005pv}
\footnote{An Euclidean version of this solution has been obtained in \cite{Iazeolla:2007wt}, and as the spin connection plays an eminent role in this solution and assuming that the action, as proposed in \cite{Boulanger:2015kfa}, is finite on this solution, we use the terminology of ``instanton solution".}
\be
\begin{split}
W' \vert_{Z=0} = W_0 - {1\over 4i} Q \o_\m{}^{\a\b}\left[(1+a^2)^2y_\a
y_\b+4(1+a^2)a_\a{}^{\ad}y_\b\yb_{\ad}+4a_\a{}^{\ad}
a_\b{}^{\bd}\yb_{\ad}\yb_{\bd}\right]-{\rm h.c.}\ ,
\label{wp}
\end{split}
\ee
subject to \eq{ss}, which serve to determine the spin connection,
and where $a_{\a\ad}$ is defined in \eq{d1}, and
\bea
a^2 &:=& a^{\a\ad}a_{\a\ad}= \frac{1-\sqrt{1-\lambda^2 x^2}}{1+\sqrt{1-\lambda^2 x^2}}\ ,
\qquad -1\le a^2\le 1\ ,
\w2
Q &=& -\frac14 (1-a^2)^2\int_{-1}^1 dt\int_{-1}^1 dt'~ 
{q(t)q(t')(1+t)(1+t')\over (1-tt' a^2)^4}\ .
\label{Q}
\eea
This gives \cite{Sezgin:2005pv} 
\bea
Q =  - {(1-a^2)^2\over 4} \sum_{p=0}^\infty 
&& \Bigg[ {2p+3\choose 2p} \left(\sqrt{1-{\nu\over 2p+1}} - \sqrt{1+{\nu\over 2p+3}} \right)^2\, a^{4p}
\\
&& - {2p+4\choose 2p+1} \left( \sqrt{1- {\nu\over 2p+3}}
-\sqrt{1+{\nu\over 2p+3}} \right)^2\, a^{4p+2} \Bigg]\ ,
\nn
\eea
exhibiting branch cuts for ${\rm Re}\, \nu\le -3$ and ${\rm Re}\, \nu\ge 1$\footnote{The unitary representations of Wigner's deformed oscillator algebra can obtained starting from the standard Fock space and factoring out ideals that depend on integer part of $(1+\nu)/2$, that is, the ideal jumps for odd values of $\nu$ \cite{Plyushchay:1997mx,Boulanger:2013naa,Vasiliev:1989re,Barabanshchikov:1996mc}. It would be interesting to examine to what extent it is possible to extend the solution to general $\nu$ properly taking into account the branch points in $Q$ at odd $\nu$.}. 
For $\nu\ll 1 $, and in the interval $-1\le a^2\le 1$, this function can be approximated by \cite{Sezgin:2005pv}
\be
Q \simeq {\nu^2 (1-a^2)^2\over 48 a^4}\left[1-{2a^2\over (1-a^2)^2}
 +{(1-a^2)^2\over 2a^2}\log {1-a^2\over 1+a^2} \right]\ .
\ee
Using \eq{ss}, one determines $\omega^{\a\b}$ and $e^{\a\ad}$ from \eq{wp}, 
while the HS Fronsdal potentials $\phi_{\mu a_1...a_s}$ vanish for $s>2$:
\be
\phi_\mu{}^{a_1...a_{s-1}} = 
\frac{\partial^{2s-2}}{\partial y^{\a_1}\cdots\partial y^{\a_{s-1}}\bar\partial \yb^{\ad_1}\cdots \bar\partial \yb^{\ad_{s-1}}} W'\vert_{Z=Y=0} =0\ , \qquad s>2\ .
\ee
Even though the HS fields vanish, it is to be noted that the solution of the metric and scalar field  constitute a solution of a highly nonlinear system of equations in which all higher derivatives play a role. One can reverse engineer a two derivative action describing the coupling of gravity to scalar field that admits the same solution \cite{Sezgin:2005pv} but such an action is clearly of limited use in the context of HS theory.

An advantage of presenting the solutions in stereographic coordinates is that it facilitates their unified description  for (A)dS. In these coordinates the solution for the scalar field is given by \eq{scalarsol} and the metric by
\bea
ds^2 &=& \frac{4\Omega^2 (d(g_1x))^2}{(1- \lambda^2 g_1^2 x^2)^2}\ ,
\w2
\Omega &=&{(1- \lambda^2  g_1^2 x^2)f_1\over 2g_1}\ ,\qquad 
g_1 = {\rm exp}~\left(\frac12 \int_1^{x^2} {f_2(t)~dt \over f_1(t)}\right)\ ,
\label{g1}
\eea
where 
\bea
&& f_1(x^2) = {2f\over h^2}\left[ 1+(1-a^2)^2Q\right]\ ,
\quad f_2(x^2)  = {16Qf\over h^2(1+h)^2}\ ,\nonumber\\
&& f(x^2)=\left[1+(1+6a^2+a^4)Q\right]^{-1}\ ,
\eea
and it is understood that the integration variable in \eq{g1} is $t=x^2$.

It is also convenient to give the result in the coordinate system defined in \eq{ads}. 
In these coordinates, the  solution takes the form \cite{Sezgin:2005hf}
\bea
 AdS_4: \quad  x^2>0 &:& ds^2\ =\ d\psi^2 +\eta^2 \sinh^2\psi\left(-d\tau^2+\cosh^2\tau~ d\Omega_2\right)\ ,
\label{adsm} \\
 && \phi = \nu \,{\rm sech}^2{\psi\over 2}\ ,
\nn\w2
 x^2<0 &:& ds^2\ =\ -d\tau^2+ \eta^2 \sin^2\tau \left(d\psi^2
 +\sinh^2\psi~ d\Omega_2\right)\ ,
 \\
 && \phi\ = \nu\, {\rm sec}^2{\tau\over 2}\ ,\
\nn\w2
dS_4: \quad x^2>0 &:& ds^2\ =\ d\psi^2 +\eta^2 \sin^2\psi\left(-d\tau^2+\cosh^2\tau~ d\Omega_2\right)\ ,
 \\
 && \phi = \nu \,{\rm sec}^2{\psi\over 2}\ ,
\nn\w2
x^2<0 &:& ds^2\ =\ -d\tau^2+ \eta^2 \sinh^2\tau \left(d\psi^2
+\sinh^2\psi~ d\Omega_2\right)\ ,
\label{dsm}\\
&& \phi\ = \nu\, {\rm sech}^2{\tau\over 2}\ ,
\nn
\eea
where we have set $|\lambda|^2=1$ and 
\be
\eta = \frac{f_1 h^2}{2} =  {1+(1-a^2)^2Q \over 1+\left(1+6a^2+a^4\right)Q} \ .
\ee
\be
AdS_4: \ \  a^2=\begin{cases} \tanh^2 \frac{\psi}{4}\ \ {\rm for}\ \ x^2>0 \\ -\tan^2 \frac{\tau}{4} 
\ \ {\rm for}\ \ x^2< 0 \end{cases}\qquad 
dS_4: \ \  a^2=\begin{cases} -\tan^2 \frac{\psi}{4}\ \ {\rm for}\ \ x^2>0 \\ \tanh^2 \frac{\tau}{4} 
\ \ {\rm for}\ \ x^2< 0 \end{cases}
\ee
In addition to the $SO(3,1)$ symmetry generated by $M'_{\a\b}$, the solution is also 
left invariant by additional transformations with rigid HS parameters
\be 
\e'\ 
=\ \sum_{\ell=0}^\infty \e'_{\ell}\
,\label{hsl21}
\ee
where the $\ell$'th level is given by \cite{Sezgin:2005pv}
\be 
\e'_\ell\ =\ \sum_{m+n=2\ell+1} \Lambda^{\a_1\dots
\a_{2m},\ad_1\dots \ad_{2n}} M'_{\a_1\a_2}\star\cdots
\star M'_{\a_{2m-1}\a_{2m}}\star \bar M'_{\ad_1\ad_2}\star\cdots\star
\bar M'_{\ad_{2n-1}\ad_{2n}}-{\rm h.c.}\ ,
\label{hsl22}\
\ee
with constant $\Lambda^{\a_1\dots\a_{2m},\ad_1\dots\ad_{2n}}$.
The full symmetry algebra is thus a higher-spin extension of $SO(3,1)\simeq
SL(2,\Comp)$, that we shall denote by
\be 
hsl(2,\Comp;\nu)\supset sl(2,\Comp)\ ,
\ee
where $sl(2,\Comp)$ is generated by $M'_{\a\b}$ and its
hermitian conjugate, and we have indicated that in general the
structure coefficients may depend on the deformation parameter
$\nu$.

Turning back to the solutions given above, a holographic and cosmological 
interpretation of \eq{adsm} has been discussed  in \cite{Sezgin:2005hf}, 
where a bouncing cosmology scenario was observed, and its comparison with 
a similar phenomenon occurring supergravities \cite{Hertog:2004rz} was made. 
In this context, it is useful to examine the behaviour of the solutions, both 
with $AdS$ and $dS$ asymptotics, near the boundary as well as distant future. 

For $|\nu|\ll 1$ and near the boundary, where $\lambda^2 x^2 \to 1$, or equivalently 
$a^2 \to 1$, the scale factors $\eta$ and $\Omega$ behaves as\footnote{The result for the limits of
$\eta$ and $\Omega$ given here correct Eqs. (4.67) and (4.68) in \cite{Sezgin:2005pv}.}
\be
\lambda^2 x^2 \to 1   \quad \implies  \quad 
\eta \to\frac{1}{1-\frac{\nu^2}{3}}\ ,\qquad \Omega \to 1\ ,
\ee
which means that the solutions are asymptotically 
maximally symmetric spacetimes with undeformed radius.

Another interesting limit to consider is $\eta\rightarrow 0$; for 
$|\nu|\ll 1$ this takes place for $a^2+1\ll 1$, that is, for
$a^2$ is close but not equal to $-1$, which corresponds to 
$\tau \to \pm \tau_{\rm crit}$ in the AdS case, and 
$\psi \to \psi_{\rm crit}$ in the dS case.
In the former case, we have\footnote{The expression for $\eta$ and $\tau_{\rm crit}$ corrects a 
factor of two in \cite{Sezgin:2005hf}.}
\be
AdS_4:\qquad  \eta \simeq \frac{\nu^2}{6} \left[ \exp \left({\frac{3}{2\nu^2}}\right) \right] 
(\tau_{\rm crit} -\tau)\ ,\qquad
\tau_{\rm crit} \simeq \sin^{-1} \left[ 2\exp \left( \frac{-3}{2\nu^2}\right) \right]\ .\ee
In Einstein frame\footnote{This is the (torsion-free) frame obtained by 
rescaling the vielbein as $e^a \to \eta^{-1} e^a$.}, it takes infinite proper time
to reach the critical surface, which means that one may interpret the future region
of the solution as a singularity free $SO(3,1)$ invariant cosmology
with a finite asymptotic scalar field, as 
\be \phi\to \phi_{\rm crit} \simeq 4\nu \exp\left(\frac{3}{\nu^2}\right)\ .\ee
In the dS case, one may instead interpret the critical limit as a domain wall
at an infinite proper space-like distance from the center of the solution.

%%%%%%%%%%%%%%%%%%%%%%%%%%%%%%%%%%%%%%%%%%%%%%%%%%%%%%%%%%%%%%%%%%%%%%%%%%%%
\subsection{Solutions of the non-minimal chiral model in Kleinian space}
%%%%%%%%%%%%%%%%%%%%%%%%%%%%%%%%%%%%%%%%%%%%%%%%%%%%%%%%%%%%%%%%%%%%%%%%%%%

In obtaining the solutions described above, symmetries on the master fields were imposed. In \cite{Iazeolla:2007wt} projection  operators were used as well.
In the case of the non-minimal chiral model\footnote{The solution can be constructed for the minimal model as well by 
working a convenient integral presentation of the projection operators.} in Kleinian space, it is possible to use projectors to build solutions with non-vanishing Weyl zero-form and HS fields. They are
\bea
B' &=& (1-P)\star\kappa\ ,\qquad S'_\a = z_\a\star P\ ,\qquad \bar S'_{\ad}\ =\ \bar z_{\ad} \star (1-2 \bar P)\ ,
\eea
where 
\bea 
P\star P&=&P\ ,\qquad \bar P \star \bar P = \bar P\ ,\qquad [P,\bar P]_\star\ =\ 0\ ,
\eea
and satisfy the conditions $\pi \circ\bar \pi (P,\bar P)  = (P,\bar P)$.  The projectors $P$ and $\bar P$ are independent.  Consider the simplest case in which
\be
 P = 2e^{-2uv}\ =\ 2e^{yby}\ , \qquad \bar P=0\ ,
\label{Pplus}
\ee
where, in terms of constant spinors $(\lambda_\a, \mu_\a)$ we have defined 
\be
u = \lambda^\a  y_\a\ ,\qquad  v= \mu^\a y_\a\  , \qquad b_{\a\b} = 2\lambda_{(\a} \mu_{\b)}\ ,
\qquad \lambda^\a \mu_\a = \frac{i}{2}\ . 
\ee
Upon the $L$-dressing, and expanding the result in $Y$-oscillators, the following component results are found in \cite{Iazeolla:2007wt}
\bea
\phi = -1\ , \qquad C_{\a_1...\a_{2s}}=0\ ,
\eea
while the anti-self-dual Weyl tensors take the form
\be
\bar C_{\ad_1\cdots\ad_{2s}}=  -2^{2s+1}(2s-1)!!\left( {h^2-1
\over h^2}\right)^s\, \bar u_{(\ad_1}\cdots \bar u_{\ad_s}\,
\bar v_{\ad_{s+1}}\cdots \bar v_{\ad_{2s})}\ ,
\label{W1}
\ee
where
\be 
\bar u_{\ad}={x^a\over \sqrt{x^2}}
\left({\bar\sigma}_a\lambda\right)_{\ad}\ , \qquad \bar v_{\ad}=
{x^a\over \sqrt{x^2}} \left({\bar\sigma}_a \mu\right)_{\ad}\ .
\label{uv}
\ee
In stereographic coordinates, the Kleinian space is covered in two charts with $0\leq h^2\leq 2$, and hence the Weyl tensors blow up in the limit $h^2\rightarrow 0$ preventing the solution from
approaching $H_{3,2}$ in this limit. In coordinate system introduced in \eq{kcs}, the metric reads $ds^2= -(dr^2+r^2 dt^2) + (1+r^2) d\Omega_2^2$, and the pre-factor in this solution reads
\be
\left( {h^2-1\over h^2}\right)^s =  2^{-s} (1-r\cos t)^s\ .
\ee
which, indeed, diverges at the boundary $r\to \infty$ as noted above. In \cite{Iazeolla:2007wt}, it was also found that 
\be 
e_\mu^a= {-2\over (h^2+2h^{-2})}\left[ (1+2h^{-2}) \delta_\mu^a +2\lambda^2 h^{-4} x_\mu x^a 
+ \frac{6\lambda^2}{h^4-4}\, (Jx)_\mu (Jx)^a\right]\ , 
\ee
where 
\be
J_{ab}=(\sigma_{ab})^{\a\b}\,b_{\a\b}\ ,\qquad J_a{}^c J_c{}^b = -\delta_a^b\ . 
\ee
For the spin connection, the result is 
\be
\begin{split} 
\omega^{\a\b} &= \frac{1}{1-4h^{-4}}\left[ \omega_0^{\a\b} -8h^{-4}
(b\,\omega_0\, b)^{\a\b}\right] + \frac{4h^{-4}}{(1-2h^{-4})(1-4h^{-4})} b^{\a\b}\, b_{\gamma\delta}\,
\omega_0^{\gamma\delta}\ ,
\label{nm2}\w2
{\bar\omega}^{\ad\bd}&= {\bar\omega}_0^{\ad\bd}
+4(1+h)^2h^{-4}\left[ -(\bar aba)^{\ad\bd} b_{\gamma\delta} +2 (\bar a b)^{\ad\gamma} (\bar ab)^{\bd\delta}
\right]\omega_{\gamma\delta}\ .
\end{split}
\ee
Note that in the last term of the second equation the full spin connection $\omega_{\gamma\delta}$ arises. 
The metric $g_{\mu\nu}=e_\mu^a e_\nu^b\,\eta_{ab}$ takes the form
\bea 
g_{\mu\nu} &=& {4\over (h^2+2h^{-2})^2}\,\Big[  (1+2h^{-2})^2 \eta_{\mu\nu} +
4h^{-4}  \left( (1-h^2) h^{-4}+ (1+2h^{-2})\right) x_\mu x_\nu 
\nn\\
&& + \frac{12}{1-4h^{-4}} \left( \frac{3 (1-h^2)}{1-4h^{-4}}+ (1+2h^{-2}\right) (Jx)_\mu (Jx)_\nu\Big]\ ,
\eea

The vierbein has potential singularities at $h^2=0$ and
$h^2=2$. The limit $h^2\rightarrow 0$ is a boundary at which 
 $e_\mu{}^a\sim h^{-2}x_\mu x^a$, \emph{i.e.} a
scale factor times a degenerate vierbein. In the limit
$h^2\rightarrow 2$ one approaches the boundary of a coordinate chart. 
Also in this limit, the vierbein becomes degenerate, 
\emph{viz.} $e_\mu{}^a\sim h^{-2}(Jx)_\mu (Jx)^a$.

%%%%%%%%%%%%%%%%%%%%%%%%%%%%%%%%%%%%%%%%%%%%%%%%%%%%%%%%%%%%%%%%%%%%%%%%%%%%%%
\subsection{Perturbative construction of domain-wall solution}
%%%%%%%%%%%%%%%%%%%%%%%%%%%%%%%%%%%%%%%%%%%%%%%%%%%%%%%%%%%%%%%%%%%%%%%%%%%%%%

As mentioned earlier, if we wish to construct a solution of the HS equations that 
has a symmetry group that includes any translation generators $P^a$, given 
that there is no known realization of these generators that would form a closed algebra 
with the full Lorentz generators $M_{\a\b}$, the symmetry conditions \eq{gsc} need 
to be imposed in terms of the undeformed generators that are bilinear in the oscillators. 
The deformation of this symmetry to accommodate nonlinear corrections can then be 
computed perturbatively in a weak field expansion scheme. This is the framework which 
was pursued in considerable detail in \cite{Sezgin:2005pv}, where the perturbative 
construction of solutions with $3$, $4$ and $6$ parameter isometry subgroups of the AdS group 
were considered. Here we shall outline the key aspects of this constructions by 
describing the example of a domain-wall solution having $ISO(2,1)$ symmetry and 
its appropriate HS extension \cite{Sezgin:2005pv}.

The $ISO(2,1)$ algebra is generated by
\be
ISO(2,1):\qquad M_{ij}\ ,\qquad P_i= (\a M_{ab}L^b +\b P_a) L^a_i\ , \qquad \a^2-\b^2=0\ ,
\ee
where $\a$ and $\b$ are real parameters and $(L_i^a,L^a)$ is a representative of the coset  $SO(3,1)/SO(2,1)$, obeying
\be 
L^a L_a\ =\ \e\ =\ \pm 1\ ,\qquad L_i^a L_a\ =\ 0\ ,\qquad
L_i^a L_{ja}\ =\ \eta_{ij}\ =\ {\rm diag}(+,+,-)\ ,
\ee
and the generators are taken form the oscillator realization of $SO(3,2)$ algebra given by
\be
 M_{ab}\ =\ -\frac18 \left[~ (\s_{ab})^{\a\b}y_\a y_\b+
 (\bar\s_{ab})^{\ad\bd}\yb_{\ad}\yb_{\bd}~\right]\ ,\qquad P_{a}\ =\
 \frac14 (\s_a)^{\a\bd}y_\a\yb_{\bd}\ ,
 \label{mab}
\ee
In particular $[P_i,P_j] = i(\b^2-\a^2) M_{ij}$ vanishes for $\alpha^2=\beta^2$, as required for $ISO(2,1)$. 

Thus, the symmetry conditions to be imposed are
\be
[M_{ij},C']=0\ ,\qquad [P_i,C']_\pi= 0\ .
\ee
As shown in \cite{Sezgin:2005pv}, these conditions are solved by
\be
C'(P) = (\mu_1+\mu_2 P) e^{4iP}\ , \qquad P:= \frac14 L^a (\sigma_a)_{\a\ad}  y^\a \yb^{\ad}\ .
\label{dws}
\ee
Denoting the  $ISO(2,1)$ transformations discussed above by $\e^\prime_{(0)}$, we can seek its nonlinear deformation by expanding 
\be
\e' = \e^\prime_{(0)}+ \e^\prime_{(1)} + \e^\prime_{(2)} + \cdots\ ,
\ee
where $\e'_{(n)}$ is constant in spacetime but may depend on $Y$ and $Z$. The symmetry condition at first order is satisfied by $C'$ given in \eq{dws}. To establish the symmetry at second order, we need to satisfy 
\be
\left( [\e^\prime_{(1)},C']_\pi + [\e^\prime_{(0)},B^\prime_{(2)}]_\pi \right)_{Z=0} =0\ ,
\label{sc5}
\ee
where $B'_{(2)}$ is obtained from the normal ordered perturbative scheme (see section 6.1 below)
to be \cite{Sezgin:2005pv}
\bea
B'_{(2)} &=& f+\tau\bar\pi f +\pi(f+\tau\bar f)^\dagger\ ,
\nn\w2
f &:=&  -z^\a \int_0^1 dt (V_\a^{\prime(1)}\star C')_{Z\to tZ}\ ,
\nn\w2
V^{\prime(1)}_\a &=& -\frac{i}{2} z_\a \int_0^1 t dt B'(-tz,\bar y) \kappa(tz,y)\ .
\eea
This condition \eq{sc5} is solved in \cite{Sezgin:2005pv} where it is found that
\be 
\e'_{(1)} =  - \int_0^1 dt \int_0^1 t'dt'\left(
{itt'\over 2}\l^{\a\b} z_\a z_\b+
\l^{\a\bd}z_\a\bar\partial_{\bd}\right)C'(-tt'z,\yb) e^{itt'yz} -h.c.\ ,
\ee
where $\lambda^{\a\b}, \lambda^{\a\bd}$ are arbitrary constant parameters. For a more detailed discussion of the procedure outlined above, see \cite{Sezgin:2005pv}.

\subsection{Other known solutions} 
The solutions described in Section 3.3 were generalized in \cite{Iazeolla:2007wt} to find new Lorentz-invariant vacuum solutions, in which in addition to the continuous parameter $\nu$, an infinite set of independent and discrete discrete parameters $\theta_k=\{0,1\}$, each turning on a Fock-space projector $P_n(u)$, were activated. Should they be proved to be gauge-inequivalent to $AdS_4$, as they seem to be, they would represent monodromies of flat but non-trivial connections $(V_\a,\bar V_{\ad})$ on ${\cal Z}$. An interesting limit of this solution arises upon setting $\nu=0$ and $(\theta_k-\theta_{k+1})^2=1$, leading to the degenerate metric 
\be
ds^2 =\frac{4 (x^a dx_a)^2}{\lambda^2 x^2 (1-\lambda^2 x^2)}\ .
\ee

The methods above can also be extended to the Prokushkin-Vasiliev theory in $D=3$, giving rise to Lorentz-invariant instanton solutions (with additional twisted sectors of the theory excited, and the characteristic extra deformation parameter $\l$ that allows to vary the mass of the scalar), as well as to the above projector vacua \cite{Iazeolla:2015tca}. 

 Finally, in \cite{Gubser:2014ysa} a different class of exact solutions was constructed by means of the gauge function method coupled with a different choice of gauge in twistor space, there referred to as axial gauge. As for the perturbative construction of solutions, it is worth mentioning the plane wave solution of \cite{Bolotin:1999fa,Vasiliev:1999ba}, whose elevation to an exact solution remains to be investigated, to our best knowledge.

%%%%%%%%%%%%%%%%%%%%%%%%%%%%%%%%%%%%%%%%%%%%%%%%%%%%%%%%%%%%%%%%%
\section{Factorization Method and Solutions}
%%%%%%%%%%%%%%%%%%%%%%%%%%%%%%%%%%%%%%%%%%%%%%%%%%%%%%%%%%%%%%%%

\subsection{The method}
%%%%%%%%%%%%%%%%%%%%%%%%%%%%%%%%%

The method developed in \cite{Iazeolla:2011cb,Iazeolla:2012nf,Iazeolla:2017vng} for finding the exact solution of Vasiliev equations also exploits the gauge function method to solve for $(B,V,W)$ in terms of $(B',V')$ from \eq{Leq} and \eq{WP}, which, in turn, are to be determined by solving \eq{P12}. It is here where the method differs from the method described above, by making the following factorized ansatz for $(B',V')$:
\bea
B'(Z,Y) &=& \nu \Psi(y,\yb)\star \kappa_y\ ,\label{B'fact}
\w2
V'_\a (Z,Y) &=& V'_\a(z;\Psi)=\sum_{n\ge 1} V_\alpha^{(n)}(z) \star \Psi^{\star n} \label{V'a}\ .
\eea
Note that this ansatz for $V'_\a$ is holomorphic in $z$, and in the following we shall refer to the solution in this form as given in \emph{holomorphic gauge} \cite{Iazeolla:2017vng}. In this section we shall consider the nonminimal model in which we recall that conditions \eq{minmod} apply.   It is shown in \cite{Iazeolla:2017vng} that this ansatz solves the fully non-linear equations \eq{P12} provided that
\be
\pi_z(V_\alpha^{(n)}(z) )=-V_\alpha^{(n)}(z)\ ,
\ee
and that 
\be
s_\alpha:=z_\alpha-2i \sum_{n\ge 1} V_\alpha^{(n)}(z) \nu^{n}\ ,
\label{se}
\ee
obeys the deformed oscillator algebra 
\be
[s_\alpha,s_\beta]_\star=-2i\epsilon_{\alpha\beta}(1-\nu \kappa_z)\ ,
\qquad \kappa_z\star s_\alpha=- s_\alpha\star \kappa_z\ . 
\label{defred}
\ee
Note that only the first power of $\nu$ in \eq{se} survives in the commutator \eq{defred}. One class of solutions is given by \cite{Iazeolla:2011cb} 
\be
\sum_{n\ge 1} V_\alpha^{(n)}(z) \nu^{n} = -\frac{i\nu}2  z_\alpha \int_{-1}^{+1} \frac{d t}{(t+1)^2}  
\exp \left(i\frac{t-1}{t+1} z_{(0)}^+ z_{(0)}^-\right){}
_1 F_1 (\tfrac12;1;b\nu \log t^2 )\ ,
\label{aa}
\ee
where the constant spinors $v_{(0)\a}^{\pm}$ are used to defined the projected oscillators 
\be
v_{(0)}^{+\alpha} v_{(0)\alpha}^{-}=1\ ,\qquad  z_{(0)}^{\pm} = v_{(0)}^{\pm\a} z_\a\ .
\label{v}
\ee
The presence of $z^+z^-$ term breaks manifest Lorentz covariance, which will be restored when we consider the field dependent gauge transformation in Section 7 that is needed to cast the results into a form that can be interpreted in terms of Fronsdal fields that obey the standard  boundary conditions.

At this point, $B'$ and $V'_\a $ are determined, with $\Psi(Y)$ representing an arbitrary initial datum. One can proceed to compute $(B,V,W')$ from \eq{P12}, \eq{Leq} and \eq{WP}. However, one needs to ensure that the star products involving $\Psi$ are well defined. The analyticity properties of the resulting $(B,V,W')$ also require special care. The strategy adopted in \cite{Iazeolla:2011cb,Iazeolla:2012nf,Iazeolla:2017vng} is to employ projection operators with well defined group theoretical origin, and easily deducible symmetry properties.  
We shall illustrate aspects of this procedure below with a relatively simple example, namely the black hole-like solution \cite{Iazeolla:2017vng} closely related to that of \cite{Didenko:2009td}, which we shall also describe in a subsequent section below.

%%%%%%%%%%%%%%%%%%%%%%%%%%%%%%%%%%%%%%%%%%%%%%%%%%%%%%%%%%
\subsection{Black hole solution}
%%%%%%%%%%%%%%%%%%%%%%%%%%%%%%%%%%%%%%%%%%%%%%%%%%%%%%%%%%

In seeking an exact solution of Vasiliev equations which has the symmetries of 4D static black hole solution that is asymptotically $AdS_4$, namely spatial rotations and time translations generated by 
\be
SO(3)\times SO(2): \qquad M_{rs}\, (r,s=1,2,3)\ , 
\ \  {\rm and}\ \  M_{05} = E= \frac14 (\sigma_0)_{\a\ad}\,y^\a\bar y^{\ad}\ .
\ee
The first instance of such a solution was found in \cite{Didenko:2009td} in a different approach that will be summarized later.
In both approaches the following projector plays an important role: 
\be
\Psi = \nu P'\ , \qquad P'\star P' = P'\ ,
\ee
where $P'$ is given by
\be
P' = 4e^{-4E}\ , 
\label{pro}
\ee
and the reality conditions dictate that 
\be
\nu= iM\ ,
\ee
with $M\in \Real$. This projector clearly has the desired symmetry property since $[M_{rs},P']_\star=0$ and $[E,P']_\star=0$. In performing the $L$-dressing, we need the result
\be
P = L^{-1} \star 4e^{-4E}\star L = 4 \exp\left (-\frac{1}{2} K_{AB} Y^AY^B\right)\ ,\label{P}
\ee
where $K_{AB}$ are the Killing parameters taking the form
\be
K_{AB} =  \begin{pmatrix} u_{\a\b} & v_{\a\bd} \\  \bar v_{\ad\b} & \bar u_{\ad\bd} \end{pmatrix}\ ,\label{KAB}
\ee
 In an $x$-dependent eigenspinors of $u_{\a\b}$, these parameters can be expressed as
\bea
u_{\a\b} &=& 2 r\, u^{+}_{(\a} u^{-}_{\b)}\ ,\qquad
v_{\a\bd} =\sqrt{1+r^2}\, \left( u^{+}_\a \bar u^{+}_{\bd} 
+ u^{-}_\a \bar u^{-}_{\bd}\right)\ ,
\nn\w2
u^{-1}_{\a\b} &=& \frac{2}{r}\, u^{+}_{(\a} u^{-}_{\b)}\ ,\qquad 
2u^{-}_{[\a}\,u^{+}_{\b]} = \e_{\a\b}\ .
\eea
The Kerr-Schild vector $k_{\a\ad}$ associated with $E$ is obtained via a projection of $v_{\a\ad}$ with the eigenspinors of $u_{\a\b}$, $\bar u_{\ad \bd}$, and can be written as\footnote{In stereographic coordinates these read \cite{Iazeolla:2017vng} $u_{\a\b}=2x^i(\sigma_{i0})_{\a\b}/(1-x^2)$ and $v_{\a\bd}= (\sigma_0)_{\a\bd} -4x_{[0} x^i (\sigma_{i]})_{\a\bd}/(1-x^2)$. }
\be
k_{\a\ad} = \frac{1}{\sqrt{1+r^2}} u^+_\a \bar u^+_{\ad} \ .
\ee
In obtaining the above results, the computation is first done in stereographic coordinates, and then coordinate change is made to go over to the spherical (global) coordinate system in which the $AdS$ metric reads $ds^2= -(1+r^2)dt^2 +(1+r^2)^{-1} dr^2 +r^2 d\Omega_2^2$.

From the point of view of the factorized Ansatz \eq{B'fact}-\eq{V'a}, the fact that $\Psi$ is proportional to a projector can be seen as a way of enforcing the Kerr-Schild property of a black-hole solution, as it effectively causes a collapse of all non-linear correction in $V'_\a,\bar V'_{\ad}$ down to the linear order, at least from the point of view of the oscillator dependence. Coupled with the gauge freedom on $S_\a$ this allows to effectively reach a gauge in which the full solution only contains first order deformations in $\nu$ \cite{Iazeolla:2011cb}. 
Another noteworthy fact is that, due to the factorized dependence on $Y$ and $Z$, one can effectively separately rotate the two oscillators by means of a factorized gauge function 
\be
g=L(x,Y)\star \widetilde L(x,Z)\ .
\label{g2}
\ee
As discussed in \cite{Iazeolla:2011cb,Iazeolla:2012nf,Iazeolla:2017vng}, turning on the second factor is useful since by choosing it appropriately one can achieve collinearity between the spin-frame $(v^+,v^-)(x)$  on $\cal Z$ (which is obtained by pointwise rotation of $v_{(0)}$ given in \eq{v} by $\widetilde L$) and the eigenspinors  $(u^{+}_{\a}, u^{-}_{\b})$ of $u_{\a\b}$ in order to remove singularities that appear in the solution for the master one-form, that are gauge artifacts. 
We note that the factor $\widetilde L(x,Z)$, being purely $Z$-dependent, does not affect $B$ and only acts non-trivially on $V_\a,\bar V_{\ad}$. 

Thus, dressing the primed fields given above by the gauge function $g$ defined in \eq{g2}, the following results have been obtained \cite{Iazeolla:2011cb}
\bea
B &=& \frac{4M}{r} \exp\left[\frac{1}{r^2}\left(\tfrac12 y^\a u_{\a\b} y^\b
+ \tfrac12 \bar y^{\ad} \bar u_{\ad\bd} \bar y^{\bd}  + i y^\a u_{\a\b} v^\b{}_{\ad} \bar y^{\ad}\right)\right]\ ,
\label{cp}\w2
S_\a &=& z_\a +8P a_\a \int_{-1}^1 \frac{dt }{(t+1+ir(t-1))^2}
\,j(t) \exp\left[\frac{i(t-1)}{ t+1+ir(t-1)}a^{+}a^{-}\right]\ ,\label{Sabh}
\w2
W' &=& W_0 + \wt L^{-1} \star d_x \wt L -\Big\{  \omega^{--}\left[ y^+y^+   +  8P  (f_1-f_2) a^{+} a^{+}\right]
\label{W'bh}\\
&& + \omega^{++}\left[y^-y^-  + 8P  (f_1-f_3) a^{-} a^{-} \right]
-2\omega^{-+} \left[ y^+ y^-  - 8 P (f_1+f_4)a^+ a^- -r (f_5+f_6) \right] \Big\}\ ,
\nn
\eea
where $y^{\pm} = u^{\a\pm} y_\a$, and similarly$(a^\pm, \omega^{++}, \omega^{--}, \omega^{-+})$ are projections of $a_\a, \omega_{\a\b}$ with $u^{\a\pm}$, and the function $j(t)$ is defined  as
\be
j(t) = -{\nu\over 4}{}_1\!
F_1\left[\frac12;2;{\nu\over 2}\log \frac 1{t^2}\right]\ .
\ee
The modified oscillators $(a_\a,\bar a_{\ad})$ are defined as 
\be
{\wt Z}_A :=  (a_\alpha,\bar a_{\dot\alpha})= Z_A + iK_A{}^B Y_B\ , \qquad [{\wt Z}_A, {\wt Z}_B] = 4i \e_{AB}\ .
\label{dos}
\ee
Furthermore, we have defined the functions
\bse\bea\label{fs}
f_1 &=& \int_{-1}^1 dt\, q(t) (t+1)\xi^3\, \exp \left[i(t-1)\xi a^{+} a^{-}\right]\ ,
\w2
f_2 &=& \int_{-1}^1  \int_{-1}^1 dt\, dt'\, \, j(t)  j(t')\, 2t\, \widetilde\xi^3\,
\exp \left[i(tt'-1)\widetilde\xi\, a^{+} a^{-}\right]\ ,
\w2
f_3 &=& \int_{-1}^1  \int_{-1}^1 dt\, dt'\, \, j(t)  j(t')\, 2t' \widetilde\xi^3\,
\exp \left[i(tt'-1)\widetilde\xi\, a^{+} a^{-}\right]\ ,
\w2
f_4 &=& \int_{-1}^1 dt\, q(t) \xi^2\, \exp \left[i(t-1)\xi a^{+} a^{-}\right]\ ,
\w2
f_5 &=& \int_{-1}^1  \int_{-1}^1 dt\, dt'\, \, j(t)  j(t')\,
 (tt'+1)\,\widetilde\xi^3 \exp\left[ itt' \widetilde \xi a^+ a^-\right]\ ,
 \w2
 f_6 &=& \int_{-1}^1  \int_{-1}^1 dt\, dt'\, \, j(t)  j(t')\,
\widetilde \xi^2 \exp\left[ i(tt'-1) \widetilde \xi a^+ a^-\right]\ ,
\eea\ese

and
\be
\xi := \frac{1}{t+1 + ir (t-1)}\ ,\qquad \widetilde\xi := \frac{1}{tt'+1 +ir (tt'-1)}\ .
\ee
The modified oscillators $a_\a$ appear naturally in $V_\a$ as a consequence of the factorized form of \eq{V'a} with $\Psi^{\star n} \propto P$, and of the fact that $z_\a\star P=a_\a P$ (analogously for $\bar a_{\ad}$ in $\bar V_{\ad}$). A consequence of the modified oscillator appearing in the solution is that $V'_\a$ does not obey the standard Vasiliev gauge $Z^A V_A=0$.

As noted earlier, this solution is closely related to that of Didenko and Vasiliev \cite{Didenko:2009td}. Indeed, the solution for $B$ is the same, while the relationship between the solutions for $(S_A,W')$ is more subtle and it will be discussed in the next section.

Note that $B$ does not depend on $Z$. Thus, $B\vert_{Z=0} = B= C(Y)$, and its  holomorphic components give the Petrov type -D Weyl tensors
\be
C_{\a_1\dots \a_{2s}} = \frac{4M}{r^{s+1}} u^{+}_{(\a_1} \dots u^{+}_{\a_s} u^{-}_{\a_{s+1}}\dots u^{-}_{\a_{2s})}\ ,
\label{ws}
\ee
and similarly for the anti-holomorphic components giving $\bar C_{\ad_1\dots \ad_{2s}}$. 
The singularity of individual Weyl tensors does not necessarily imply a physical singularity in HS gravity for the following reasons. At the level of the master fields, $r=0$ also appears as the only point at which $V_\a$, as well as $W'$, acquire a pole on a plane in $\cal Z\times \cal Y$ defined by $a_\a|_{r=0}=z_\a +i(\s_0\yb)_\a=0$, due to the zero at $t=-1$ that the denominator of the integrand in \eq{Sabh} develops at $r=0$. The master-field curvature is however given by $B\star\kappa$ and one can argue that at the master-field level, which is the only sensible way to look at such solution in the strong-field region, $B$ remains, in fact, regular at $r=0$. Qualitatively this can be understood as follows. $r$ appears in \eq{cp} as the parameter of a delta sequence: away from the origin one has a smooth Gaussian function, approaching a Dirac delta function on $\cal Y$ as $r$ goes to zero \cite{Iazeolla:2011cb}. However, unlike the delta function on a commutative space, the delta function in noncommutative twistor space, thought of as a symbol for an element of a star product algebra, is smooth. 
Indeed, it is possible to show \cite{Iazeolla:2011cb} that by changing ordering prescription one can map the delta function to a regular element, and the smoothness of such change of basis manifests itself in the fact that the solution of the deformed oscillator problem obtained in the new ordering can be mapped back smoothly to the solution above. In this sense, the singularity in $r=0$ may be an artifact of the ordering choice for the infinite-dimensional symmetry algebra governing the Vasiliev system.

In order to extract the $x$-dependence of even just the spin-$2$ component of $W'$, one still needs to evaluate the complicated parametric integrals in \eq{fs}. However, as the Weyl-tensors take the simple form in x-space given in \eq{ws}, we expect that a suitable gauge transformation exists that will give the metric in the standard Kerr-Schild form, namely $g_{\mu\nu} = g_{\mu\nu}^{AdS} + 2Mk_\mu k_\nu/r$. 

Finally, let us note that the basic black hole-like solution reviewed above has a generalization in which infinite set of projection operators $P_n$ and twisted projection operators $\widetilde P_n$ are introduced\footnote{Even though $\widetilde P'_n \star\widetilde P'_n = P'_n$, we refer to $\widetilde P'$ as twisted projector to emphasize the fact that it is related to the projector $P'_n$ by the relation $\widetilde P'_n = P'_n \star \kappa_y$.}. The twisted projectors $\widetilde P_n$ are invariant under $SO(3)\times SO(2)$ discussed above while the projectors $P_n$ are invariant under $SO(3)$ and the $(B',S'_A)$ sector of the solution takes the form \cite{Iazeolla:2011cb,Iazeolla:2012nf,Iazeolla:2017vng}
\bea 
B' & = & \sum_{n=\pm1, \pm2,...}\left(\nu_n \widetilde P'_n + \widetilde\nu_n P'_n \right)\ , 
\qquad \widetilde P'_n = P'_n \star \kappa_y\ ,
\label{ban}\\[5pt]
S'_A & = & z_A -2i\sum_{n=\pm1,\pm2,...} \left(V_{n,A}\star P'_n + \widetilde V_{n,A}\star \widetilde P'_n\right)  \ ,
\label{wSprime}
\eea
with
\bea 
P'_{n}(E) & = &  2(-1)^{n-\ft{1+\ve}2}\,\oint_{C(\ve)} \frac{d\eta}{2\pi i}\,\left(\frac{\eta+1}{\eta-1}\right)^{n}\,e^{-4\eta E} \ ,\qquad \varepsilon := n/|n|\ ,
\label{pn}\w2
\widetilde P'_n (E) & := & P'_n(E)\star\k_y \ = \ 4\pi (-)^{n-\ft{1+\ve}2}\,\oint_{C(\ve)} \frac{d\eta}{2\pi i}\,\left(\frac{\eta+1}{\eta-1}\right)^{n}\,\d^2(y-i\eta \s_0\yb) \ ,
\label{ptn}
\eea
where the contour integrals are performed around a small contour $C(\ve)$ encircling $\ve$. The expressions for $(V_{n,A}, \widetilde V_{n,A})$, and their  $g=L\star \widetilde L$ dressing can be found in \cite{Iazeolla:2017vng}. It turns out that the solutions with only $\nu_n$ parameters switched on correspond to black hole like solutions, of which the case summarized above arises for $n=1$.  Solutions with only nonvanishing $\widetilde\nu_n$ parameters correspond to massless particle modes, and surprisingly black hole modes as well entering from second order onwards in a perturbative treatment of the solution; for a detailed description of this phenomenon see \cite{Iazeolla:2017vng,Carlo}. Note that the $SO(3)$ invariant projectors are associated with spin-$0$ modes. To extend this construction to spin-$s$ particle modes, one needs the spin-$s$ generalization of the projectors $P'_{n}$ discussed above. A particular presentation of such projectors can be found in \cite{Iazeolla:2008ix}.

\subsection{Other known solutions} 

By means of the same factorized Ansatz \eq{B'fact}-\eq{V'a} black-hole-like solutions with biaxial symmetry have also been found \cite{Iazeolla:2011cb,Iazeolla:2012nf,Sundell:2016mxc}, some of which being candidate HS generalizations of the Kerr black hole \cite{Sundell:2016mxc}. The separation of variables in holomorphic gauge was also instrumental to finding solutions with $\mathfrak g_6$ isometries of cosmological interest, that we shall present in a forthcoming paper \cite{cosmo}.

%%%%%%%%%%%%%%%%%%%%%%%%%%%%%%%%%%%%%%%%%%%%%%%%%%%%%%%%%%%%%%%%%%%%%%%
\section{Direct Method and the Didenko-Vasiliev Solution} 
%%%%%%%%%%%%%%%%%%%%%%%%%%%%%%%%%%%%%%%%%%%%%%%%%%%%%%%%%%%%%%%%%%%%%%

While, as we have seen so far, the gauge function method is in general of great help in constructing exact solutions, it is sometimes possible to attack the equations directly, by virtue of some other simplifying Ansatz or gauge condition. We shall generically refer here to any method which does not rely on the use of gauge function as direct method. One such solution has been found so far in this way by Didenko and Vasiliev \cite{Didenko:2009td}, which has nonvanishing HS fields, and which contains the Schwarzschild black hole solution in the spin 2 sector.

Indeed,  motivated by the phenomenon that solutions of Einstein equations that can be put in Kerr-Schild form solve the linearized as well as the nonlinear form of the equations, the Authors of \cite{Didenko:2009td} thought of an Ansatz that would generalize some distinctive features of black hole solutions in gravity. First, it is based on an AdS timelike Killing vector, in the sense that the Weyl tensor will be a function of some element $K_{AB}$ as \eq{KAB}. More precisely, if $f(K)$ satisfies the Killing vector equation, a proper ansatz for a solution of the linearized twisted adjoint equation will be given by $f(K) \star \k_y$ . Second, they chose $f(K)$ in such a way that the Ansatz linearize the Vasiliev equations. For the latter purpose it is important that the function $f(K)$ is a projector --- in fact, a Fock-space vacuum projector that coincides with \eq{P}. Such choice in particular reduces equations \eq{S2} and \eq{S3} to two copies of the 3D (anti)holomorphic deformed oscillator problem that arises in Prokushkin-Vasiliev HS theory in 3D \cite{ProkVas} in terms of the oscillators in \eq{dos}\footnote{In this section we shall use the conventions of \cite{Didenko:2009td} which differ form ours.}.

The ansatz \cite{Didenko:2009td}
\be
B= M P\star \kappa_{y}\ ,\qquad S_\a= z_\a + P f_\a (a| x)\ , \qquad \bar S_{\dot\a}= \bar z_{\dot\b} + P \bar f_{\dot\a}(\bar a|x)\ ,
\ee
where $M$ is a constant and $(f_\a,\bar f_{\dot\a})$ are functions to be determined, indeed reduces the 
equations \eq{S2} and \eq{S3} to two deformed oscillator problems in terms of the latter functions. A specific gauge choice on the $(f_\a,\bar f_{\dot\a})$, while bringing about a further breaking of the manifest Lorentz covariance, effectively linearizes their equations, which are then solved by means of standard perturbative methods \ref{standardscheme}. A further ansatz \cite{Didenko:2009td}
\be
W=W_0 +  P \left[g (a|x) + \bar g(\bar a|x) \right]\ ,
\ee
where $g$ is another function to be determined is then employed to deal with the 
remaining equations that involve $W$, namely \eq{B1}, \eq{B2} and \eq{S1}.  
The resulting exact solution is given by \cite{Didenko:2009td}
\bse\bea
B&=&\frac{4M}{r} \exp\left[\frac{1}{r^2}\left(\ft12 y^\a u_{\a\b} y^\b
+ \ft12 \bar y^{\ad} \bar u_{\ad\bd} \bar y^{\bd}  + y^\a u_{\a\b} v^\b{}_{\ad} \bar y^{\ad}\right)\right]\ ,
\w2
S^+ &=& z^+ + MP \frac{a^+}{r} \int_0^1 dt \exp \left(\frac{t}{r} a^{+} a^{-}\right)\ ,
\ \ {\rm idem}\ \ \bar S^+\ ,
\w2
S^- &=& z^- \ ,
\ \ {\rm idem}\ \ \bar S^-\ ,
\w2
W &=& W_0 + \frac{1}{2r} M P \left[ d\tau^{--} 
a_\alpha^+  a_\beta^+ \int_0^1dt (1-t) \exp \left(\frac{t}{r} a^{+} a^{-}\right) +h.c. \right] 
\nn\\
&& + \frac{1}{8r} MP  \left[ u_{\a\b}\, \omega_0^{\a\b} + h.c. + 2 (v_{\a\ad} + k_{\a\ad})\, e_0^{\a\ad}\right]\ ,
\eea\label{DV}\ese
where
\be
\begin{split}
\tau_{\a\b} &= \frac{u_{\a\b}}{r}\ ,\qquad \tau^{--}=u^{\a-}u^{\a-}\t_{\a\b}\ ,\qquad z^{\pm}=u^{\a\pm}z_\a\ ,\qquad a^{\pm} = u^{\a\pm} a_\a\ ,
\w2
d\tau_{\a\b} &= -\frac{1}{r} \omega_{0(\a}{}^\gamma\, u_{\b)\gamma} 
+\frac{1}{r} e_0^{\gamma\dot\gamma} \left(\e_{\gamma(\alpha}\,v_{\b)\dot\gamma} 
+\frac{1}{2r^3} v_{\dot\gamma}{}^\delta u_{\a\b} u_{\gamma\delta}\right)\ .
\end{split}
\ee
Note also that $S_A$ does not satisfy the Vasiliev gauge $S(Z,Y)\vert_{Z=0}=0$, and that $W$ above has not been redefined as in \eq{WP}. Nonetheless, it has been noted in \cite{Didenko:2009td}  that a HS transformation of the form $ \delta W= D_0 \e^{(1)}$ with 
$ \e^{(1)} = \left(-\frac12 \int_0^1 dt z^\a S_\a\vert_{z\to tz} + h.c.  + f (Y|x) \right)$ and arbitrary $f(Y|x)$ maps  $W$ to $W^{\rm phys}$ given by
\be
W^{\rm phys} = \frac{4M}{r} e_0^{\a\ad} k_{\a\ad} \exp \left(-\frac12 k_{\b\bd} y^\b \bar y^{\dot\b}\right)\ , \label{Wphys}
\ee
whose spin 2 component gives the frame field and the associated metric 
\be
e_\mu^a = e_\mu{}^{a (AdS)} + \frac{M}{r} k_\mu k^a\ ,\qquad g_{\mu\nu} = g^{\rm AdS}_{\mu\nu} +\frac{2M}{r} k_\mu k_\nu\ ,
\ee
where $k_\mu = e_0^{\a\ad}\,k_{\a\ad}$. This is the metric of a black hole of mass $M$ in $AdS_4$ in Kerr-Schild form. The terminology of black hole in HS context requires caution as discussed in the introduction. In addition to the $SO(3)\times SO(2)$, the solution summarized above has been shown to also have $1/4$ of the ${\cal N}=2$ supersymmetric HS symmetry of the model, and their infinite dimensional extension thereof \cite{Didenko:2009td}. 

The solution \eq{DV} differs from \eq{cp}-\eq{W'bh} both in the form of the internal connection $(V^\pm,\bar V^\pm)$ and in that of the gauge field generating functions. As for the internal connection, the difference can be ascribed to the two choices employed in \cite{Iazeolla:2011cb} and \cite{Didenko:2009td} for solving the deformed oscillator problem (referred to  as ``symmetric'' and ``most asymmetric'', respectively, in \cite{Iazeolla:2011cb,Iazeolla:2012nf}). One can show that the resulting internal connections can be connected via a gauge transformation (see \cite{Iazeolla:2011cb}), although the small or large nature of this transformation is yet to be investigated. The comparison of $W'$ in \eq{W'bh} and $W$ in \eq{DV} is technically more complicated, as the two differ also by the shift of the Lorentz connection \eq{WP}, and it will be postponed to a future work, but we note that having the same $B$ identical in both solutions strongly suggests that the physical gauge fields should be equivalent, and in particular equivalent to \eq{Wphys}.

It is worth mentioning that even by working without the gauge function method, with a specific choice of gauge the Didenko-Vasiliev solution can be simplified in such a way that the $W$ connection is reduced to the vacuum one $W_0$. This simplification was studied in \cite{Bourdier:2014lya}, along with the embedding of the solution in the ${\cal N}=2$ and ${\cal N}=4$ supersymmetric extensions of the bosonic Vasiliev equations.  

%%%%%%%%%%%%%%%%%%%%%%%%%%%%%%%%%%%%%%%%%%%%%%%%%%%%%%%%%%
\section{Perturbative Expansion of Vasiliev Equations }
\label{standardscheme}
%%%%%%%%%%%%%%%%%%%%%%%%%%%%%%%%%%%%%%%%%%%%%%%%%%%%%%%%%%

In this Section, we shall summarize the standard perturbative expansion of the Vasiliev equations, benefiting from \cite{Vasiliev:1999ba,Sezgin:1998gg,Sezgin:2002ru,Sezgin:2011hq,Iazeolla:2011cb,Boulanger:2015ova} (for more recent treatments, see \cite{Didenko:2015cwv,Vasiliev:2017cae}).

In the normal order, defined by the star product formula \eq{star},
the inner Klein operators become real analytic in $Y$ and $Z$ space,
\emph{viz.}
\be
\kappa=\kappa_y\star \kappa_z= \exp(iy^\alpha z_\alpha)\ ,\qquad
\bar\kappa=\kappa_{\bar y}\star \kappa_{\bar z}= 
\exp(i\bar y^{\dot\alpha} \bar z_{\dot \alpha})\ .
\ee
Assuming that the full field configurations are real-analytic on
${\cal Z}_4$ for generic points in ${\cal X}_4$, one may thus 
impose initial conditions 
\be
B|_{Z=0}=C\ ,\qquad W_\mu|_{Z=0}=a_\mu\ ,
\ee
where the $x$-dependence is understood. 
In order to compute the $Z$-dependence of the fields
one may choose  $V_A|_{C=0}$, which is trivial flat connection,
to vanish, and a homotopy contractor for the de Rham differential
on ${\cal Z}_4$, which entails imposing a gauge condition 
on $V_A$.
One may then solve the constraints on $D_A B$, 
$F_{AB}$ and $F_{A\mu}$ on ${\cal Z}_4$ in a
perturbative expansion in $C$.
This procedure gets increasingly complex with increasing order 
in the expansion, which schematically can be written as 
\bea
B &=&\sum_{n\geqslant 1}B^{(n)}\ ,\qquad
B^{(1)}(C)\equiv C\ ,
\w2
V&=& \sum_{n\geqslant 1}V^{(n)}\ ,
\w2
W &=& \sum_{n\geqslant 0 }W^{(n)}(a_\mu)\ ,\qquad
W^{(0)}(a_\mu)\equiv a_\mu\ ,
\eea
where $B^{(n)}$, $V^{(n)}$ and $W^{(n)}(a_\mu)$ are $n$-linear functionals
in $C$, and $W^{(n)}(a_\mu)$ is a linear functional in $a_\mu$.
These quantities, which are constructed using the homotopy contractor on ${\cal Z}_4$,
depend on $Z$, and are real-analytic in ${\cal Y}_4\times {\cal Z}_4$
provided that $C$ and $a_\mu$ are real analytic in $Y$-space 
and all star products arising along the perturbative expansion  
are well-defined.
As for the remaining equations, that is, that $F_{\mu\nu}=0$ and $D_\mu B=0$, 
it follows from the Bianchi identities that they are perturbatively equivalent
to $F_{\mu\nu}|_{Z=0}=0$ and $D_\mu B|_{Z=0}=0$, which form a perturbatively 
defined Cartan integrable system on ${\cal X}_4$ for $C$ and $a_\mu$.

To Lorentz covariantize, one imposes
\be W'|_{Z=0}=w \ ,\ee
with $W'$ from \eq{WP}, that is 
\be 
a_\mu=w_\mu+\frac1{4i}\left.(\omega^{\a\b}_\mu 
M_{\a\b}+\bar\omega^{\dot\a\dot\b}_\mu  \bar M_{\dot\a\dot\b})\right|_{Z=0}\ ,
\ee
where $w$ does not contain any component field proportional to 
$y_{\a}y_{\b}$ and $\bar y_{\dot\a} \bar y_{\dot\b}$ in view of \eq{ss}.
Upon substituting the above relation into $W^{(n)}(a_\mu;C,\dots,C)$,
it follows from the manifest Lorentz covariance that the dependence 
of $F_{\mu\nu}|_{Z=0}$ and $D_\mu B|_{Z=0}$ on the Lorentz connection 
arises only via the Lorentz covariant 
derivative $\nabla$ and the Riemann two-form $(r^{\alpha\beta},r^{\dot\a\dot\b})$. 
Thus, the resulting equations in spacetime take the form \cite{Iazeolla:2011cb}
\be 
\nabla w +w\star w+\frac1{4i} \left( r^{\a\b}y_\a y_\b 
+ h.c.\right)  
+i\sum_{\tiny\begin{array}{c}n_1+n_2\geqslant 1
\nn\\ 
n_{1,2}\geqslant 0 \end{array}}\left(
r^{\alpha\beta}
V^{(n_1)}_\alpha\star V^{(n_2)}_\beta + h.c.\right)\Big|_{Z=0}
\ee
\be 
+\sum_{\tiny\begin{array}{c}n_1+n_2\geqslant 1\\ n_{1,2}\geqslant 0
\end{array}}
W^{(n_1)}(w)\star  W^{(n_2)}(w)
\Big|_{Z=0}=0\ ,
\label{v1}
\ee
\be 
\nabla C \ +  
\sum_{\tiny \begin{array}{c}n_1+n_2\geqslant 1
\\ n_1\geqslant 0,\ n_2\geqslant 1\end{array}}[W^{(n_1)}(w_\mu),B^{(n_2)}]_\pi\Big|_{Z=0}=0\ ,
\label{v2}
\ee
where 
\bea
\nabla w &:=& d_x w + [\omega^{(0)}, w]_\star\ ,\quad \nabla C := d_x C +[\omega^{(0)},C]_\star 
\ , \quad 
\omega^{(0)} = \frac{1}{4i} \omega^{\a\b} M^{(0)}_{\a\b}\ ,
\w2
r^{\a\b} &:=& d\omega^{\a\b}-\omega^{\a\gamma}\wedge \omega_{\gamma}{}^\b\ ,\qquad
{\bar r}^{\dot\a\dot\b}:= d{\bar \omega}^{\dot\a\dot\b}-{\bar \omega}^{\dot\a\dot\gamma}
\wedge {\bar \omega}_{\dot\gamma}{}^{\dot\b}\ .
\label{rbr}
\eea

Alternatively, in order to stress the perturbation expansion around the maximally 
symmetric background, including the spin-two fluctuations, it is more convenient
to work in terms of the original one-form $a_\mu$. 
The perturbative expansion up to 3rd order in Weyl curvatures reads 
\bea
da &=& a\star a+\cV(a,a,C)+ \cV^2(a,a,C,C) + \cO(C^3)\ ,
\label{zh1}
\w2
dC &=& a\star C -C \star \pi(a) -\cU(a,C,C) + \cO(C^3)\ ,
\label{zh2}
\eea
where $(\cV,\cV^2,\cU)$ are functionals that can be determined from \eq{v1} and \eq{v2}.
One next assumes that the homotopy contraction in $Z$-space is performed 
such that
\be
z^A V_A=0\ ,
\ee
which we refer to as the Vasiliev gauge, and expands
\be
a = W_0 + a_1+a_2+ \cdots\ ,\qquad   C= C_1+C_2+ \cdots\ .
\ee
where $W_0$ is the maximally symmetric background; $a_1$ a
and $C_1$ are linearized fields; $a_n$ and $C_n$ are
$n$th order fluctuations.
The resulting linearized field equations on $X$-space 
provides an unfolded description of a dynamical scalar field 
\be
\phi=C_1\mid_{Y=0}\ ,
\ee
and a tower of spin-$s$ Fronsdal fields 
\be
\phi_{a(s)}=\left( e^{\mu a}_{0} \left((\sigma_a)^{\alpha\dot\alpha} 
\frac{\partial^2}{\partial y^\alpha \partial {\bar y}^{\dot\alpha}}\right)^{s-1} a_{1,\mu}\right)\Bigg\vert_{Y=0=Z}\ ,
\ee
where we use the convention that repeated indices are symmetrized.
Computing the functional $\cV(W_0,W_0,C)$, the linearized 
unfolded system is given by \cite{Vasiliev:1999ba}
\bea
D_0 a_1  &=& -\frac{i}{4} H^{\a\b} \del_{\a}\del_{\b} C_1 (y,0) -
\frac{i}{4} \bar H^{\ad\bd} \bar{\del}_{\ad}\bar{\del}_{\bd} C_1 (0,\yb) 
\label{fe}\w2 
D_0 C_1 &=&  0\ ,
\label{sfe}
\eea
where  
\bea
D_0 a_1 &:=& d_x a_1 + \{ W_0, a_1 \}_\star\ , 
\qquad D_0 C_1 := d_x C +[W_0,C_1]_\star\ ,
\w2
h^{\a\ad} &:=& e_0{}^{\a\ad}\ ,\qquad H^{\a\b} :=  h^{\a\dot\c}\wedge h^{\b}{}_{\dot\c} \ ,\qquad 
 \bar H^{\ad\bd} := \bar h^{\ad\c}\wedge \bar h^{\bd}{}_{\c}\ . 
\eea
The oscillator expansion of \eq{fe} furnishes the definition 
of the spin$-s$ Weyl tensors, and gives the field equations 
for spin-$s$ fields which remarkably do not contain higher 
than second order derivatives, and indeed they are the well 
known Fronsdal equations for massless spin $s$-fields in $AdS$. 
As for \eq{sfe}, its oscillator expansion gives the AdS massless 
scalar field equation in unfolded form.

Perturbative expansion around $AdS$ at second order is rather 
complicated but still manageable.  
Schematically they take the form 
\bse
\bea
&& D_0 a_2 -\cV(W_0,W_0,C_2) = a_1\star a_1 +2 \cV(W_0,a_1, C_1) 
+ \cV(W_0,W_0,C_1, C_1)\ ,
\label{zz1}\w2
&& D_0 C_2 = [a_1,C_1]_\pi  + \cU(W_0,C_1,C_1)\ .
\label{zz2}
\eea
\ese
Even though these terms have been known in the form of parametric 
integrals for some time, their detailed structure and consequences 
for the three point functions  were considered much later in 
\cite{Giombi:2009wh,Giombi:2010vg}, where the field equation for 
the scalar field was examined, and used for computing the three-point amplitude for spins  
$0-s_1-s_2$.
If $s_1\ne s_2$, only the first source term in \eq{zz2} contributes and gives a finite result,
in agreement with the boundary CFT prediction. 
However, if $s_1=s_2$, only the second source 
term in \eq{zz2} contributes and gives a divergent result \cite{Giombi:2009wh,Giombi:2010vg}. 
This divergence was confirmed later \cite{Boulanger:2015ova,Skvortsov:2015lja} 
in the three point amplitude for spins $s-0-0$, resulting from the 
last term in \eq{zz1}. 
Soon after, it was shown that a suitable redefinition of the 
master zero-form cures this problem \cite{Vasiliev:2016xui,Vasiliev:2017cae}, 
as has been also confirmed with the computation of relevant 
three-point amplitudes \cite{Sezgin:2017jgm,Didenko:2017lsn}. 
A similar redefinition in the one-form sector has also been 
determined so that the divergence problem arising in the last 
term in the first equation above has also been removed 
\cite{Gelfond:2017wrh}. 

In determining the higher order terms in the perturbative expansions 
of Vasiliev equations, it remains to be established in general what
field redefinitions are allowed in choosing the appropriate basis 
for the description of the physical fields.
In wrestling with this problem, the remarkable simplicity of 
the holographic duals of this highly nonlinear and seemingly 
very complicated interactions may provide a handle by means of their 
holographic reconstruction. Such reconstruction has been achieved for the three 
and certain four-point interactions \cite{Bekaert:2014cea,Bekaert:2015tva,Sleight:2016dba}. 
Putting aside the analysis and interpretation of the nonlocalities \cite{Bekaert:2015tva,Sleight:2017pcz,Ponomarev:2017qab}, which are present, and nonetheless in accordance with holography by construction, the issue of how to extract helpful hints from them with regard to the nature of the allowed field redefinitions in perturbative analysis of Vasiliev equation remains to be seen.

Of course, ultimately it would be desirable to have a direct formulation 
of the principles that govern nonlocal interactions, based on 
the combined boundary conditions in twistor space as well as 
spacetime, as we shall comment further below.

%%%%%%%%%%%%%%%%%%%%%%%%%%%%%%%%%%%%%%%%%%%%%%%%%%%%%%%%%%%%%%
\section{A Proposal for an Alternative Perturbation Scheme}
%%%%%%%%%%%%%%%%%%%%%%%%%%%%%%%%%%%%%%%%%%%%%%%%%%%%%%%%%%%%%%

In what follows, we shall show that for physically relevant initial data 
$\Psi(Y)$ given by particle and black hole-like states, the solutions obtained 
using the factorization method can be mapped to the Vasiliev gauge used in the 
normal ordered perturbation scheme at the linearized level.
Whether the Vasiliev gauge is compatible with an asymptotic description in terms of
Fronsdal fields to all orders in perturbation theory, or if it has to be modified,
possibly together with a redefinition of the zero-form initial data, is an open 
problem.
In finding the proper boundary conditions in both spacetime and twistor
space it may turn out to be necessary to require finiteness of a set of 
classical observables involving integration over these spaces.

One can define formally the aforementioned map to all orders of classical 
perturbation theory by applying a gauge function
\be G=L\star H\ ,\ee
to the holomorphic gauge solution space, where $H=1+\sum_{n\geqslant 1} H^{(n)}$ is a field dependent gauge function to be fixed as to impose the Vasiliev gauge in normal ordering.
To begin with, let us consider fluctuations around $AdS$ for which $\Psi(Y)$ 
consists of particle states, focusing, for concreteness, to the case of scalar particle states worked out in \cite{Iazeolla:2017vng}.
Upon switching on the gauge function $L$, the field $V_\a^{(L)}:=L^{-1}\star 
V'_\a\star L$ develops non-analyticities in the form of poles in 
$Y$-space in the particle sector \cite{Iazeolla:2017vng}.
Applying $H$ removes these poles and expresses the results in terms of Fronsdal fields (at the same time restoring the manifest Lorentz covariance, as we shall see), at least in the leading order. This can be see as follows. We want to obtain $V^{(G)}_\a$ such that
\bea 
z^\a \wV^{(G)}_\a +\mbox{h.c.} =  
z^+\wV^{( G)-} - z^-\wV^{(G)+}  +\mbox{h.c.}\ = \ 0 \ . 
\label{VFgauge} 
\eea 
The leading order gauge transformation reads
\bea  
\wV_\a^{(G)(1)} = \wV^{(L)(1)}_\a +\partial_\a  H^{(1)} \ .
\label{VFtransf}
\eea
Contracting by $z^\a$ and using the fact that by definition $Z^A \wV^{(G)}_A=0$, one finds 
\bea  
H^{(1)} = -\left(\frac{1}{z^\b\partial_\b}z^A \wV^{(L)(1)}_A +\mbox{h.c.}\right)\ .
\eea
In particular, activating only the scalar ground state (and its negative-energy counterpart) via the parameters $\widetilde \nu_{\pm 1}$ in the projector ansatz for $B$ (see \eq{ban}), this was computed in \cite{Iazeolla:2017vng} with the result 
\bea 
H^{(1)} \ = \ \left.-\frac{i}{4}\,\frac{1-x^2}{1-2i x_0 + x^2}\,
\frac{1}{\ty^+\ty^-}\,\frac{\ty^+z^- + \ty^- z^+}{\widetilde u}\, 
\left(e^{i\widetilde u}-1\right)\right|_{\eta=+1} + \mbox{idem}|_{\eta=-1}\ ,
\eea 
where we have set $b=\widetilde \nu_{1}={\widetilde\nu}_{-1}=1$, and 
\be 
\widetilde u  :=  \ty^\a z_\a  =  \ty^+z^- - \ty^- z^+\ ,
\qquad \ty_\a = y_\a + M_\a{}^{\bd}(x,\eta)\yb_{\bd}\ .
\ee
The matrix $ M_\a{}^{\bd}(x,\eta)$ can be found in Section 5.2.1 of \cite{Iazeolla:2017vng}.
This result for $H^{(1)}$ is regular in ${\cal Z}$ but has a pole in ${\cal Y}$.
It follows that 
\bea 
\wV_\a^{(G)(1)} &=& -\frac{1}{2}\,\frac{1-x^2}{1-2i x_0 + x^2}\,\frac{z_\a}{\widetilde u}\left[e^{i\widetilde u}-\frac{e^{i\widetilde u}-1}{i\widetilde u}\right]+ 
\mbox{idem}|_{\eta=-1}\ ,
\label{fv}\w2
B(Y|x) &=& \frac{4(1-x^2)}{1-2i x_0 + x^2}\,\left.e^{iy^\a M_\a{}^{\bd} \yb_{\bd}} 
\right|_{\eta=+1} +  \mbox{idem}|_{\eta=-1}\ .
\eea
We observe that $\wV_\a^{(G)(1)}$ is now real-analytic \emph{everywhere} on ${\cal C}$.
Furthermore, it was shown in \cite{Iazeolla:2017vng} that the above expressions for $(B,V_\a)$ lead to the relation
\be 
\wV^{(G)(1)}_\a = z_\a\int_0^1 dt \,t\,B(-tz,\bar{y}) \,e^{ity^\a z_\a}   \,
\label{V1}
\ee
in agreement with the result obtained in the standard perturbative analysis of Vasiliev equations at leading order. The emergence of $z_\a$ in \eq{fv} shows that manifest Lorentz covariance is restored, in comparison with the expression for $\wV_\a^{(L)(1)}$. 
The prefactor in \eq{fv} is a consequence of the fact that we are considering the lowest mode alone in the solution for Fronsdal equation the scalar field, as opposed to summing all the full set of modes.

Expressing the exact solutions obtained in holomorphic gauge in terms of Fronsdal fields amounts to setting up an alternative perturbation scheme in which one constructs the higher orders of the gauge function $H^{(n)}$ subject to \emph{dual boundary conditions}, that is, to conditions restricting the both the twistor-space dependence of the master fields and their spacetime asymptotic behaviour. 
Indeed, one requires that after having switched on $H$, the master fields have symbols
in normal order that are real-analytic at $Y=Z=0$, and symbols in Weyl order that 
belong to a (associative) star-product subalgebra with well-defined classical observables
defined by traces over twistor space and integrations over cycles in spacetime.
The proposal is that this problem admits a non-trivial solutions, and that it fixes 
$H^{(n)}$ up to residual small HS gauge transformations and the initial data for the 
master zero-form $B$ to all orders in classical perturbation theory.
It would be interesting to see whether these type of field redefinitions are 
related to those recently proposed by Vasiliev in order to obtain a quasi-local perturbation theory in terms of Fronsdal fields \cite{Vasiliev:2016xui,Gelfond:2017wrh}. 
An important related issue is whether the gauge function $G$  
is large in the sense that it affects the values of the HS zero-form charges, 
which are special types of classical observables given by traces over 
twistor space defining zero-forms in spacetime that are de Rham closed \cite{Sezgin:2005pv,Sezgin:2011hq,Bonezzi:2017vha}.
%

%%%%%%%%%%%%%%%%%%%%%%%%

A nontrivial test of the factorization approach is first to show that the solution is finite after performing the higher order $H^{(n)}$ gauge transformations, and second, to show that the resulting $n$-point correlators are in agreement with the result expected from holography. The corrections beyond the leading order remain to be determined, while the computation of correlators has been performed in which the second order solution in standard perturbative scheme has been used. It has been shown that a naive computation of $B^{(2)}$ in the standard perturbative scheme leads to divergences \cite{Giombi:2009wh,Boulanger:2015ova}, and later it was shown that these divergences can be removed by a suitable redefinition of $C(Y|x)$ \cite{Vasiliev:2016xui,Vasiliev:2017cae}. Whether there exists a principle based on any notion of quasi-locality in spacetime that governs the nature of such redefinitions to all orders in perturbation is not known, to our best knowledge. 

An advantage of the factorization method is that here we start from a full solution
to Vasiliev's theory, defined as a classical field theory on the product of spacetime
and twistor space (not referring a priori to the conventional perturbative approach). 
This provides a convenient framework for the description of the solutions with
particles fluctuating around nontrivial backgrounds. 
The key principle here is that linear superposition principle holds for the 
zero-form initial data $\Psi(Y)$. 
For example, if we want to describe the solution to Vasiliev equations for 
particles propagating around BH solution of Section 4, one simply takes 
$ \Psi= \nu_n P_n + \widetilde \nu_n \widetilde P_n$. 
The exact solution for the combined system is obtained in this way, but in 
small fluctuations of a particle propagating in a fixed and exact black hole 
background, one may treat the parameter $\widetilde \nu_n$ and $\nu_n$ as 
small and large, respectively. 
A very interesting open problem is thus to combine this scheme with the
aforementioned proposal for dual boundary conditions in order to
work out new types of generating functions for HS amplitudes.

%%%%%%%%%%%%%%%%%%%%%%%%%%%%%%%%%%%%%%%%%%%%%%%%%%%%%%%%%%%%
\section{Aspects of Higher Spin Geometry}
%%%%%%%%%%%%%%%%%%%%%%%%%%%%%%%%%%%%%%%%%%%%%%%%%%%%%%%%%%%%

HS theory has been mostly studied  at the level of the field equations and 
in terms of locally defined quantities in ordinary spacetimes or twistor 
space extensions thereof. 
It is clearly desirable to develop a globally defined framework for the 
classical theory, in order to provide geometric interpretations of the exact 
solutions, and shed further light on the important issue of
the choice of boundary conditions on the master fields on the total
space required for Vasiliev's
equations to produce physically meaningful anti-holographic duals of 
boundary conformal field theories. 
An attempt at a global description of HS theory was made in \cite{Sezgin:2011hq} where bundle structures based on different choices of structure 
groups, soldering forms and classical observables were considered\footnote{
We refer the reader to \cite{Sezgin:2011hq} for considerable amount of details
albeit using a a considerably different notation.}.
Here, we shall highlight a particularly interesting choice of structure group, 
and the resulting infinite dimensional coset description involving tensorial
coordinates, and associated generalized frame field and resulting metrics. 

\subsection{Structure group}
%%%%%%%%%%%%%%%%%%%%%%%%%%%%%%%%%%%%%%%%
%
Noting that Vasiliev equations are Cartan integrable in arbitrary 
number of commuting dimensions, yet without changing the local 
degrees of freedom associated to the zero-forms (more on this below),
we consider the formulation of Vasiliev's theory in terms of the 
master fields $(W,B,S_\a,\bar S_{\dot\a})$ thought of as horizontal forms 
on a noncommutative fibered space ${\cal C}$ with eight-dimensional fibers
given by ${\cal Y}_4\times{\cal Z}_4$ and base given by an infinite-dimensional commuting real manifold $\cal M$, consisting of charts coordinatized by $X^{M}$.
In each chart, the one-form field $W=dX^{M} W_{M}$ thus 
takes its values in the HS Lie algebra
\be 
\widehat{\mathfrak{hs}}(4) := \left\{\widehat P(Y,Z) ~:~\tau(\widehat P) = (\widehat P)^\dagger~=~-\widehat P\right\}\ ,  
\label{hs4}
\ee
and the deformed oscillators $(S_\a,\bar S_{\dot\a})$ and the deformation field $B$
are thought of as zero-forms on ${\cal M}$, valued in representations of the HS Lie algebra as described in section 2, where the $\tau$-map is also defined.
The extended equations of motion are given by Eqs. \eq{B1}--\eq{S3}, with the 
only difference being that the manifold $\chi_4$ is replaced by
${\cal {M}}$. 

In order to define the theory globally on ${\cal M}$, we glue together 
the locally defined master fields using transition functions from 
a structure group, which by its definition is generated by a structure 
algebra ${\mathfrak h}$ given by a subalgebra of $\mathfrak{hs}(4)$. 
One interesting choice is \cite{Sezgin:2011hq}
\be
{\mathfrak h} = \widehat{\mathfrak{hs}}_{+}(4)\oplus {\mathfrak{sl}}(2,\Comp)\ ,\qquad \widehat{\mathfrak{hs}}_+(4) := \frac12( 1+\pi ) \widehat{\mathfrak{hs}}(4)\ ,
\ee
where ${\mathfrak{sl}}(2,\Comp)$ is the algebra of canonical Lorentz
transformations.
The corresponding connection, also referred
to as the generalized Lorentz connection, is given by 
\be 
\Omega :=  W^+\oplus  \omega \ ,\qquad 
W^\pm := \frac12 (1\pm\pi) W'\ ,
\ee
where $\omega$ is the canonical Lorentz connection, 
while the projection 
\be {\cal E} := W^-\ ,\ee
is assumed to form a section\footnote{It is worth noting that in \cite{Sezgin:2011hq}, the form ${\cal E}$ was considered to be a
soldering form on a manifold $\underline{\cal M}$ with tangent space isomorphic to the coset $\widehat{\mathfrak{hs}}(4)/\widehat{\mathfrak{hs}}_{+}(4)$ and containing ${\cal M}$ as a submanifold. We have simplified the geometrical framework here by formulating the system directly on ${\cal M}$.}.
The Vasiliev equations and gauge transformations 
in terms of these master fields are spelled out in \cite{Sezgin:2011hq}.

\subsection{Soldering mechanism}
%%%%%%%%%%%%%%%%%%%%%%%%%%%%%%%%%%%%%%%%%%%%%%%%%%%%%%%%%%%%%%%%
%
The horizontal differential algebra on ${\cal C}$, which is quasi-free in the sense 
that the curvature constraints are cartan integrable modulo zero-form
constraints, can be projected to a free horizontal differential algebra on 
the reduced total space ${\cal C}|_{Z=0}$.
To this end, one first solves the constraints on ${\cal Z}$ given 
initial data 
\be 
w(Y|X) \equiv  dX^M w_M  = 
dX^{M}\,\widehat W'_{M}\Big|_{Z=0}\ ,
\qquad  C(Y|X) = B\Big|_{Z=0}\ ,
\ee
where $w(Y|X)$ belongs to the reduced HS algebra 
\be
\mathfrak{hs}(4) := \widehat{\mathfrak{hs}}(4)\Big|_{Z=0}\ ,
\ee
and $C(Y|X)$ belongs to its twisted adjoint representation. 
Next, defining
\be
 \Gamma \oplus \omega:=\Omega\Big|_{Z=0} \in \mathfrak{hs}_+(4) \oplus \mathfrak{sl}(2,\Comp)
\ ,\qquad
E: = \cE\Big|_{Z=0}\in \mathfrak{hs}(4)\,/\,\mathfrak{hs}_+(4)
\ ,
\ee
where $\mathfrak{hs}_+(4):= \frac12(1+\pi)\mathfrak{hs}(4)$,
we assume that ${\cal M}$ is soldered by $E$, that is,
the tangent space of ${{\cal M}}$ is assumed to be 
identified with the coset ${\mathfrak{hs}}(4)\,/\,{\mathfrak{hs}}_{+}(4)$
via $E$.
Thus, expanding 
\be E= dX^M E_M^A P_A\ ,\qquad \pi(P_A) = -P_A\ ,
\ee
where $P_{A}$ is a basis for the $\pi$-odd elements of ${\mathfrak{hs}}(4)$,
and denoting the inverse of the frame field $E_{M}{}^{ A}$ by
$E^{M}{}_{ A}$, the local translations with gauge parameters
$\xi=\xi^{A}P_{A}$ can be identified 
by usual means \cite{Sezgin:2011hq} as infinitesimal diffeomorphisms
generated by globally defined vector fields $\vec \xi=\xi^{A} \cE^{ M}{}_{ A}\vec\partial_{M}$ combined with local generalized Lorentz transformations with parameters $(\imath_{\vec \xi}\, \Gamma,\imath_{\vec \xi}\,\omega)$.

\subsection{Elimination of tensorial coordinates}
%%%%%%%%%%%%%%%%%%%%%%%%%%%%%%%%%%%%%%%%%%%%%%%%%%%%%%%%%%%%%%%%

The framework described above can be used to write the zero-form constraint  on ${\cM}\times \{Z=0\}$, up to leading order in $C$, as
\be
D_A(\C,\o) C + \{P_A,C\}_\star=0\ ,
\label{Ceq}
\ee
where $D(\C,\o)=\nabla+{\rm ad}_\C$ with $\nabla$ 
representing the Lorentz covariant derivative.
This constraint can be analyzed by decomposing $P_{A}=\left\{ P_{A_\ell}\right\}_{\ell=0}^\infty$ into levels of increasing tensorial rank, \emph{viz.}
\be
P_{A_\ell} = \left\{P_{a(2\ell+1),b(2k)}\right\}_{k=0}^\ell~=~\left\{M_{\{a_1b_1}\star \cdots \star M_{a_{2k} b_{2k}}\star P_{a_{2k+1}}\star \cdots \star P_{a_{2\ell+1}\}}\right\}_{k=0}^{\ell}\ ,
\ee
where $(M_{ab},P_a)$ are the generators of $SO(3,2)$ and $P_{a(2\ell+1),b(2k)}$ is a Lorentz tensor of type $(2\ell+1,2k)$.
The zeroth level of the zero-form constraint reads
\be 
D_a(\C, \o) C +\{P_a,C\}_\star = 0\ , \qquad a=1,...,4\ ,
\ee
where the translation generator is given by the twistor relation
$P_a = (\s_a)^{\a\ad}y_\a \yb_{\ad}/4$,
which implies that $C_{\a(n+2s),\ad(n)}$ is identifiable as the $n$th order symmetrized covariant vectorial derivative of the primary spin-$s$ Weyl tensor $C_{\a(2s)}$ \cite{Sezgin:2002ru}. 
On the other hand, the $\ell$th level of the zero-form constraint implies
\be 
D_{a(2\ell+1)}(\C, \o) C+\{P_{a(2\ell+1)},C\}_\star = 0\ ,
\ee
where the higher translation generator is now given by the enveloping formula
\be P_{a(2\ell+1)} = P_{\{a_1}\star\cdots P_{a_{2\ell+1}\}}\ .
\ee
As a result, the tensorial derivatives $\nabla_{a(2\ell+1)}(\C, \o)$ factorize 
on-shell into multiple vectorial derivatives; for example, the tensorial 
derivative of the physical scalar $\phi=C|_{Y=0}$ factorizes into
\be 
D_{a(2\ell+1)}(\C, \o)\phi~\propto~ D_{\{a_1}(\C, \o)\cdots
D_{a_{2\ell+1}\}}(\C, \o)\phi\ .
\ee
It follows that no new strictly local degrees of freedom are introduced due to the presence of the tensorial coordinates of {${{\cal M}}$.

\subsection{Generalized metrics}
%%%%%%%%%%%%%%%%%%%%%%%%%%%%%%%%%%%%%%%%%%%%%%%%%%%%%%%%%%%%%
%
Taking traces of $\star$-products of generalized vielbeins 
$\cE=dX^{M} \cE_{M}$ and adjoint operators on twistor space, 
one can construct structure group invariants that are 
tensor fields on $\cal M$ \cite{Sezgin:2011hq}; in particular,
we may consider symmetric rank-$r$ tensor fields 
\be 
G_{(r)} := dX^{{\underline M}_1}\cdots  dX^{{\underline M}_r} 
{\rm Tr}\left[{\cal K}\star \left( \cE_{{M}_1}\star {\cal V}^{k_1,\bar k_1}_{\l_1,\bar \l_1}\right) \star \cdots\star 
\left( \cE_{{ M}_r}\star {\cal V}^{k_r,\bar k_r}_{\l_r,\bar \l_r}\right) \right]\ ,
\label{metric}
\ee
where ${\cal K} \in \left\{1,\kappa\bar\kappa\right\}$ and 
\be
{\cal V}^{k,\bar k}_{\l,\bar \l}  := \left\{ \exp_\star \left[i( \l^\a S_{\a}+\bar\l^{\ad} \bar S_{\ad})\right] \right\} \star ( B\star\kappa)^{\star k}\star  (B\star {\bar\kappa})^{\star \bar k}\ ,\qquad k,\bar k\in\mathbb N\ ,
\label{vertex}
\ee
where $(\l^\a,\bar\l^{\ad})$ are auxiliary twistor variables, and
\be {\rm Tr}\left[f(Z,Y)\right] := \int_{{\cal Y}\times {\cal Z}} \frac{d^4Y d^4Z}{(2\pi)^4}\,  f(Z,Y)\ .\ee
Given a (compact) $p$-cycle $[\S]$ in the homology of ${\cal M}$, 
one can consider the formally defined minimum
\be 
{\cal A}_{\rm min}[\Sigma,G_{(r)}]~:=~ \min_{\Sigma'\in [\Sigma]} {\cal A}[\Sigma',G_{(r)}]\ ,
\ee
of the area functional
\be  
{\cal A}[\S',G_{(r)}] := \int_{\Sigma'}d^p\sigma \left(
\e^{m_1[p]}\cdots \e^{m_r[p]} \underbrace{(f^\ast G)_{m_{1}\dots m_{r}}\cdots 
(f^\ast G)_{m_{1}\dots m_{r}}}_{\tiny \mbox{$p$ times}}\right)^{1/r}\ ,
\ee
defined using the totally anti-symmetric tensor density
$\e^{m_1\dots m_p}$ and induced metric
\be 
(f^\ast G)_{m_1\dots m_r} = \partial_{m_1} X^{{\underline M}_1}\cdots\partial_{m_r} X^{{\underline M}_s} G_{{\underline M}_1\dots {\underline M}_r}\ ,
\ee
on $\S'$ equipped with local coordinates $\s^m$.
The area functional is structure group invariant, and 
its minimum, if well-defined, is ${\rm Diff}({\cal M})$ 
invariant, hence serving as a classical observable for the 
HS theory.
Clearly, the dressings by the vertex operator-like operators
result in a large number of inequivalent metrics for each 
$r$, but this is not a novelty in HS theory, as 
it is possible to consider similar dressings 
of the Einstein frame metric in ordinary gravity.
The tensorial calculus pertinent to examining the 
variational principles for $r=2$ is well understood, whereas
for $r>2$ it remains to be investigated further.
In particular, one may ask whether there exists any principle for
singling out a preferred metric (possibly of rank two),
using, for example, calibrations based on abelian $p$-forms of the 
type that will be discussed below.
The application of these ideas to a geometrical characterization of the exact  solutions of HS theory remains to be investigated.

\subsection{Abelian $p$-form charges}
%%%%%%%%%%%%%%%%%%%%%%%%%%%%%%%%%%%%%%%%%%%%%%%%%%%%%%%%%%%%%%%%%%%

Another type of intrinsically defined classical observables facilitated 
by the introduction of soldering one-forms are the charges of on shell 
closed abelian $p$-forms, \emph{viz.}
\be
{\cal Q}[\Sigma,H] = \oint_{\S} H(\cE,B,S_\a, \bar S_{\ad}, r_{\a\b},\bar r_{\ad\bd})\ ,
\label{calQ}
\ee
where $\S$ are closed $p$-cycles in ${\cal M}$ or open $p$-cycles with  suitable boundary conditions, and $H$ are globally defined differential forms that are cohomologically nontrivial on shell, namely
$dH = 0$ and $\delta_{\L}H = 0$ ($\Lambda\in \mathfrak{h}$),  and $H$ is not globally exact. 
We also recall that $r_{\a\b}$ is defined in \eq{rbr}. 
The abelian $p$-forms were considered in \cite{Sezgin:2011hq}
\be  
H_{[p]} = {\rm Tr} \left[ (\cE\star \cE+r^{(S)+})^{\star (p/2)}\star \kappa\right]\ ,\qquad p=2,4,6,...
\label{Hp}
\ee
where
\be  
r^{(S)+} = \frac{1}{8i} (1+\pi) r^{\a\b} S_\a \star S_\b  -h.c.
\label{geodefs}
\ee
The conservation of these charges can be checked 
directly by replacing the exterior derivative by the structure
group covariant derivative $D(\Omega)$ inside the trace and
using the fact that $D(\Omega)( \cE\star \cE+ r^{(S)+})=0$ on shell.

\subsection{On-shell actions} 

Actions have been proposed that imply the Vasiliev equations \cite{Boulanger:2011dd,Boulanger:2015kfa} upon applying the variational principle. 
Their on-shell evaluations involve subtleties stemming from their 
global formulation and the crucial role played by the boundary conditions. 
Thus, it remains unclear whether these actions vanish on-shell. 
Nonetheless, one can still employ the abelian $p$-form charges 
discussed above as off-shell topological deformations. 
To this end, one set of candidates that have been considered are 
\cite{Sezgin:2011hq}
\be
\begin{split}
S_{\rm top}[\S_2] &= {\rm Re} \left\{ \tau_2 \oint_{\S_2} {\rm Tr} \left[\kappa\star R\right]\right\}\,
\w2
S_{\rm top}[\S_4] &={\rm Re}\left\{ \oint_{\S_4}
{\rm Tr}\left[\kappa\star \left(  \tau_4 R\star R+ \tilde \tau_4\left((\cE\star \cE+r^{(S)+})\star
R+\frac12 (\cE\star \cE+ r^{(S)+})^{\star 2}\right)\right)\right]\right\}\ ,
\label{tvo}
\end{split}
\ee
where $\tau_2$, $\tau_4$ and $\tilde \tau_4$ are complex constants, $\S_{2,4}$ are submanifolds of ${\cal M}$ (where Lagrange multipliers vanish), and 
\be
R := \nabla W^{+} + W^{+}\star W^{+} + \frac{1}{4i} (\omega^{\a\b} M^{(0)}_{\a\b} + {\bar\omega}^{\ad\bd} \bar M^{(0)}_{\ad\bd})\ ,
\ee
which is the curvature of $\Omega$.
Using the field equation \cite{Sezgin:2011hq}
\be
R+\cE \star \cE + r^{(S)+} =0\ ,
\ee
we can express $S_{\rm top}[\S_4]$ on-shell as
\be
S_{\rm top}[\S_4]~\approx~ {\rm Re}\left\{(\t_4-\frac12 \tilde \tau_4)\oint_{\S_4}H_{[4]}\right\}\ ,
\label{ampl}
\ee
where $H_{[4]}$ given in \eq{Hp}. 
Infinities arise from the integration over $\Sigma_4$ as well as twistor space.
Assuming that the divergence from the AdS vacuum has a definite reality
property, such that it can be removed by choosing $\tau_4$ appropriately, 
one is left with an integral over perturbations that may in principle be finite modulo a prescription for integration contours; for related discussions, see \cite{Giombi:2010vg,Colombo:2010fu}. 
Provided that one considers perturbations corresponding to boundary sources, 
it would be natural to interpret $S_{\rm top}[\S_4]$ as the generating functional 
for the boundary correlation functions. 

One may also construct topological two-forms given on-shell by 
\be 
S_{\rm top}[\S_2]~\approx ~ - {\rm Re}\left\{\tau_2 \oint_{\S_2}H_{[2]}\right\}\ ,
\ee
with $H_{[2]}$ given in \eq{Hp}.
One application of this surface operator is to wrap $\Sigma_2$ around a point-like defect or singularity such as the center of the rotationally symmetric and static solution of \cite{Didenko:2009td}. In this case, the leading order contribution, which comes from the AdS vacuum,  is a divergent integral over twistor space. If the divergence has a definite reality property, one can cancel it by choosing $\tau_2$ appropriately. One would then be left with an integral over the perturbations. As the latter involve nontrivial functions in twistor space, the integral may be finite. It would be interesting to seek an interpretation of the resulting value of $S_{\rm top}[\S_2]$ as some form of entropy of the black hole solutions reviewed above.

An alternative framework in which certain $2$- and $4$-forms in $x$-space, referred to as the Lagrangian forms, are introduced was proposed in \cite{Vasiliev:2015mka}, where their possible application for the computation of HS invariant charges and generating functional for the boundary correlations functions were discussed; see also \cite{Didenko:2015pjo} for the computation of such HS charges on black-hole solutions at the first order in the deformation parameters. Asymptotic charges in HS theories have also been described in \cite{Barnich:2005bn,Campoleoni:2017vds,Campoleoni:2016uwr}. Their relation to the HS charges\footnote{It is worth noting that the HS charges described here, as well as those based on the above mentioned Lagrangian 2-form and evaluated in \cite{Didenko:2015pjo}, present significant differences with respect to the asymptotic charges. For instance, the first ones are closed and gauge invariant everywhere in spacetime, which makes them proper classical observables also in the strong-coupling regions. Crucial for this to happen is that they contain contributions from all spins. On the other hand, each spin-$s$ asymptotic charge is conserved per se, but is crucially defined only via an integration over a two-cycle at spatial infinity under the hypotesis of $AdS$ asymptotics, i.e., only in the weak-coupling region. Indeed, it was shown in \cite{Didenko:2015pjo} that the HS charge based on the Lagrangian 2-form can be written as a combination of contributions from each spin-$s$ field integrated over any two-cycle. Each of the latter contributions gives rise to a separately conserved spin-$s$ charge only when the latter two-cycle is pushed to spatial infinity. Moreover, the asymptotic charges depend for their definition on the existence of asymptotic symmetries, whereas the HS charges, being purely master field constructs, are conserved even without assuming any symmetry. As stressed in \cite{Didenko:2015pjo}, this can be seen as a consequence of the non-local expansion in derivatives of the physical fields that is naturally contained in the master fields on-shell.} as well as their evaluation on certain exact solutions will be discussed elsewhere \cite{Carlo}.

\subsection*{Acknowledgements}
%%%%%%%%%%%%%%%%%%%%%%%%%%%%%%%%%%%%%%%

We thank David De Filippi, Gary Gibbons, Dmitry Ponomarev, 
Eugene Skvortsov and Xi Yin for helpful discussions in the 
course the writing of this review. 
The work of CI is supported in part by the Russian Science Foundation grant 14-42-00047 in association with the Lebedev Physical Institute in Moscow. 
The work of ES is supported in part by NSF grant PHY-1214344. 
The work of P.S. is supported by Fondecyt Regular grants N$^{\rm o}$ 1140296 and N$^{\rm o}$ 1151107 and Conicyt grant DPI 2014-0115.

\newpage

%%%%%%%%%%%%%%%%%%%%%%%%%%%%%%%%%%%%%%%%%%%%%%%%%%%%%%%%%%%%%%%%%%%%%%

\end{document}